\documentclass[12pt]{article}
\usepackage{amsmath}
\usepackage{cite}
\usepackage[toc,page]{appendix}
\usepackage{bbold}
\usepackage{amssymb}
\usepackage{epsfig,graphicx}
\usepackage{cancel}
\usepackage{graphicx}
\usepackage{capt-of}
\usepackage{float}
\usepackage{lipsum}
\usepackage{subcaption}
\usepackage{tikz}
\usepackage{float}
\usepackage{diagbox}
\usepackage{multirow}

\def\be{\begin{equation}}
\def\ee{\end{equation}}
\def\bea{\begin{eqnarray}}
\def\eea{\end{eqnarray}}

\newcommand{\normord}[1]{:\mathrel{#1}:}

\begin{document}
\begin{center}

{\large \bf Non-commutative probability insights into the double-scaling limit SYK model with constant perturbations:  moments, cumulants and $q$-independence}\vskip 0.5cm

{\large \bf Shuang Wu}\footnote{wushuang@itp.ac.cn}
\\[0.1cm]
{\it CAS Key Laboratory of Theoretical Physics, Institute of Theoretical Physics,
\\Chinese Academy of Sciences, Beĳing 100190, China}
\\[0.1cm]
\end{center}

\vskip 0.5cm
\centerline{\large \bf Abstract}
\vskip 0.2cm  
Extending the results of \cite{Wu}, we study the double-scaling limit SYK (DSSYK) model with an additional diagonal matrix with a fixed number $c$ of nonzero constant entries $\theta$. This constant diagonal term can be rewritten in terms of Majorana fermion products. Its specific formula depends on the value of $c$. We find exact expressions for the moments of this model. More importantly, by proposing a moment-cumulant relation, we reinterpret the effect of introducing a constant term in the context of non-commutative probability theory. This gives rise to a $\tilde{q}$ dependent mixture of independences within the moment formula. The parameter $\tilde{q}$, derived from the $q$-Ornstein-Uhlenbeck ($q$-OU) process, controls this transformation. It interpolates between classical independence ($\tilde{q}=1$) and Boolean independence ($\tilde{q}=0$). The underlying combinatorial structures of this model provide the non-commutative probability connections. Additionally, we explore the potential relation between these connections and their gravitational path integral counterparts.

\section{Introduction}
The Sachdev-Ye-Kitaev (SYK) model \cite{Sachdev1,Sachdev2,Kitaev} is known as a solvable toy model in AdS/CFT holography, which is often
used to study the quantum gravity problems such as black hole thermodynamics and the information paradox \cite{Kitaev2,Cotler,Chen}. It is dual to the Jackiw-Teitelboim (JT) gravity \cite{Jackiw,Teitelboim} in the low energy limit and the connection is via the Schwarzian action \cite{Maldacena}. One of the most important features of the SYK model is that it is maximally chaotic \cite{Maldacena2}, \i.e. it saturates the bound on chaos\cite{Maldacena3} and it's late time behaviour can be captured by random matrix statistics\cite{Saad1}.
In this paper, we are particularly interested in the double scaling limit SYK (DSSYK) model, which becomes exactly solvable over the whole energy spectrum in this limit. The ensemble averaged moments of this model can be efficiently mapped to a combinatorial counting problem involving chord diagrams \cite{Erdos, Berkooz1}. Initially, these chord diagrams were only used as a tool for obtaining the eigenvalue distribution. More recently, however, research \cite{Lin, Berkooz3} has shown that the DSSYK model has its own gravitational picture, with chord diagrams serving as a suitable framework for describing its bulk theory. In \cite{Susskind,Narovlansky} a relation of the DSSYK model to de Sitter space has also been proposed. The DSSYK has a distinct chaotic behaviour in comparison to the original SYK model. The operator growth generated by the DSSYK Hamiltonian has a vanishing scrambling time at infinite temperature, as demonstrated in \cite{Bhattacharjee}, it is called hyperfast scrambling in \cite{Susskind1}. This suggests that the DSSYK model is a holographic model of de Sitter space rather than an anti-de Sitter model. Nevertheless, if the low-energy and $q\rightarrow 1$ limits are carefully taken, many results for the original SYK model can be recovered, including the maximum chaos exponent.  Further details can be found in \cite{Berkooz2}. 

The objective of this study is to investigate the DSSYK model with the inclusion of an additional constant perturbation source. This is a continuation of our previous work \cite{Wu}. The original concept was developed from random matrix theory. It is well known that spiked random matrices that have been deformed by either additive or multiplicative perturbations, exhibit a sharp phase transition in which the outliers (largest eigenvalues) are extracted from the bulk of the eigenvalue density \cite{Johnstone,Baik,Peche}. The corresponding eigenvectors of the deformed model also have an analogous localization/delocalization transition \cite{Benaych}. Thus, we assume that, given the parallels between the SYK model and random matrix theory, we should apply a similar approach and introduce an additional perturbation source to assess its impact on the properties of the SYK model. The simplest case is that of adding a constant term as a perturbation. This idea is, at least to some extent, correct, as we observe a phase transition at the eigenvalue level. However, this transition is more complex than that observed in the spiked random matrix case. This paper only focuses on the eigenvalue distribution, as the behaviour of the eigenvectors of the perturbed DSSYK model is more complicated and beyond the scope of this study. Nevertheless, through the moment calculation, we have identified a potential connection with non-commutative probability theory, which may provide insight into the eigenvectors problem. This is once again inspired by random matrix theory. For spiked random matrices, the limiting behaviour of the eigenvectors associated with the outliers is determined by the free subordination function, as described in \cite{Capitaine1, Biane}. This function  characterises a Markov transition kernel, which projects the old basis of the extra source to the new basis of the perturbed model. In this paper, we identify a similar Markov transition, namely the conditional expectation associated with the q-OU process discussed in Section \ref{sec:4}. The investigation of the eigenvectors of our perturbed DSSYK model will be deferred to future work. Furthermore, it is important to highlight that free probability theory has recently been widely applied in many-body physics, including the generalized eigenstate thermalization hypothesis (ETH) \cite{Pappalardi1, Pappalardi2} and mesoscopic models of quantum transport \cite{Hruza, Bernard}. The fundamental concept is the definition of a local freeness notion for the probed operators, based on their moment-cumulant relation. The concept of local freeness is defined by the dependence of the cumulants on the index. The index in question may be used to represent different positions or different energy levels, depending on the system and the operator under study.
Consequently, the connection established in this study between the DSSYK model and non-commutative probability theory may prove useful in studying the dynamics of its time-dependent correlation functions.

Considering the Sachdev-Ye-Kitaev (SYK) model model plus a diagonal matrix with $c$ constant non-zero entries $\theta$, denoted as $\theta$:
\begin{align}
H:&=H_{\text{SYK}}+\theta D_c\nonumber\\
&=i^{p(p-1)/2}\sum_{1\leq i_1<i_2<\dots<i_p\leq N} J_{i_1,i_2,\dots,i_p} \psi_{i_1}\psi_{i_2}\dots\psi_{i_p}+\theta\text{diag}{(1,\ldots,1,0,\ldots,0)}\label{hamiltonian}
\end{align}
the first part above  represents the standard SYK Hamiltonian with $p$ $\psi_{j}$  interacting Majorana fermions among $N$
Majorana fermions satisfying $\lbrace\psi_{i},\psi_{j}\rbrace=2\delta_{i,j}$. The factor $i$ in front of \eqref{hamiltonian} is introduced to keep $H_{\text{SYK}}$ Hermitian.
$J_{i_1,i_2,\dots,i_p}$ are random coupling satisfying
\begin{align}
\langle J_{i_1,i_2,\dots,i_p}^2\rangle_{J}=\frac{1}{\binom{N}{p}}
\label{Jsclae}
\end{align}
where $\langle \cdots\rangle_{J}$ represents the ensemble average over the random couplings $J$ \footnote{Note that the variance \eqref{Jsclae} is not a common choice. In the double-scaling limit \eqref{double-scaled}, it is different from the usual SYK conventions only by a constant factor\cite{Berkooz2}. This is a normalization factor which corresponds to the number of terms under the sum over the index sets in \eqref{hamiltonian} and is chosen to keep the spectrum of order $1$.}. We choose this scaling to simplify the calculation, with \eqref{Jsclae}, we normalize the trace of SYK Hamiltonian $ \langle \text{tr} H_{\text{SYK}}^2 \rangle_J  =1$\footnote{trace operator in this paper is normalized as $\text{tr} \mathbb{1}=1$.}
. Therefore, moments calculation can be reduced to a combinatorial counting problem \cite{Berkooz1,Berkooz2}. In the following, we are interested in studying the distribution of eigenvalues of \eqref{hamiltonian}, in the double-scaling limit \cite{Garcia1,Berkooz1,Erdos}
\begin{equation}
N \rightarrow \infty, \quad \lambda=\frac{2 p^{2}}{N}=\text { fixed. }\label{double-scaled}
\end{equation}
First of all,  define $\Psi_{\alpha}=i^{p(p-1)/2}\prod_{i=1}^p\psi_{i}$ as the product of $p$ Majorana fermions, $\alpha$ represents an index set
as $\alpha=\left\lbrace i_{1},i_{2}\cdots i_{p}\right\rbrace $ which satisfy
\begin{align}
\Psi_{\alpha}^{2}=\mathbb{1}, \quad \Psi_{\alpha} \Psi_{\beta}=(-1)^{c_{\alpha \beta}+p} \Psi_{\beta} \Psi_{\alpha}\label{gammarel}
\end{align}
where $c_{\alpha \beta}=\vert \alpha\cap\beta\vert$ is the  number of common indices in the sets $\alpha$ and $\beta$. In the double-scaling limit, $c_{\alpha \beta}$ follows a Poisson distribution with mean $p^2/N$\cite{Berkooz1}. 
In the following we assume $p$  and $N$  always even. In addition, we introduce a new variable $q$ as an average factor to count the exchange between any two $\Psi$ in this double scaling limit \eqref{double-scaled}
\begin{align}
q:=\sum_{c_{\alpha \beta}=0}^{\infty}\frac{(p^2/N)^{c_{\alpha \beta}}}{c_{\alpha \beta}!}(-1)^{c_{\alpha \beta}}\exp(-p^2/N)=\exp(-\lambda)
\label{q}
\end{align}
therefore $q\in[0,1]$. We also assume that no three sets of fermions can have common elements simultaneously, $\vert\alpha\cap\beta\cap\gamma\vert=0$.
In this limit, the ensemble-averaging moments of the SYK model are mapped to an enumeration problem, evaluated by a transfer matrix $T$, which is simply a sum of creation and annihilation operators on the chord Hilbert space. See section \ref{sec:2} for a detailed review of the functioning of the transfer matrix method in the double-scaled SYK model. Thus, it is equivalent to assuming that no three chords in a diagram intersect at the same point.

$D_c$ can be rewritten as a linear combination of tensor products of Pauli matrix $\sigma_3$ and identity matrix $\sigma_0$ :
\begin{align}
D_c:&=\text{diag}{(\underbrace{1,1,\ldots,1}_\text{$c= 2^{N/2-k}$},0,0,\ldots,0)}\nonumber\\
&=2^{-k}\sum_{ i_1, i_2,\ldots,i_k\in\lbrace 0,3\rbrace}\underbrace{\sigma_{i_1}\otimes \dots\otimes \sigma_{i_k}}_\text{$\sharp =k$}\otimes \underbrace{\sigma_{0}\otimes\dots\otimes \sigma_{0}}_\text{$\sharp =N/2-k$}
\end{align}
We denote $r=2^{-k}$ as the proportion of nonzero entries in the diagonal of $D_c$.

Since the Majorana fermions can be also written as a tensor product of Pauli matrices, 
\begin{align}
\psi_l=\underbrace{\sigma_{1}\otimes \dots\otimes \sigma_{1}}_\text{$\sharp =N/2-\lfloor l/2\rfloor$}\otimes\sigma_{x}\otimes \underbrace{\sigma_{0}\otimes\dots\otimes \sigma_{0}}_\text{$\sharp =\lfloor  l/2\rfloor  -1$}\quad\text{with}\quad \sigma_{x}=\left\{\begin{array}{ll}\sigma_{2} \quad & \text{$l$ even} \\ \sigma_{3} \quad & \text{$l$ odd}\end{array}\right.
\label{pauli}
\end{align}
where $\lfloor l/2\rfloor$ denotes the integer part of $l/2$. We can further write this perturbation diagonal term in terms of Majorana fermions
\small
\begin{align}
D_{2^{(N-2k)/2}}=\frac{1}{2^k}\sum_{m=0}^{k-1}(-i)^m\sum_{l_1=0}^{k-2}\sum_{l_2=l_1+1}^{k-2}\dots\sum_{l_m=l_{m-1}+1}^{k-2}\prod_{n=0}^{m-1}\psi_{N-2l_{m-n}-1}\psi_{N-2l_{m-n}}(\mathbb{1}_{2^{N/2}}+(-i)^{N/2}\prod_{j=1}^N\psi_{j})
\label{pert}
\end{align}
\normalsize
For examples, the half, the quarter and the one-sixth of the diagonal elements are  nonzero:
\small
\begin{align*}
D_{2^{(N-2)/2}}&=\frac{1}{2}\left(\mathbb{1}_{2^{N/2}}+\sigma_{3}\otimes\underbrace{\sigma_{0}\otimes \dots\otimes \sigma_{0}}_\text{$\sharp =N-1$}\right)\\
&=\frac{1}{2}\left(\mathbb{1}_{2^{ N/2}}+(-i)^{N/2}\psi_1\psi_2\dots\psi_{N}\right)
\end{align*} 
\begin{align*}
D_{2^{(N-4)/2}}=\frac{1}{4}\left(\mathbb{1}_{2^{N/2}}+(-i)^{N/2}\psi_1\psi_2\dots\psi_{N}+(-i)^{(N-2)/2}\psi_1\psi_2\dots\psi_{N-2}-i\psi_{N-1} \psi_N\right)
\end{align*}
\begin{align*}
D_{2^{(N-6)/2}}&=\\
&\frac{1}{8}(\mathbb{1}_{2^{N/2}}+(-i)^{N/2}\prod_{j=1}^N \psi_j+(-i)^{(N-2)/2}\prod_{j=1}^{N-2}\psi_j+(-i)^{(N-4)/2}\prod_{j=1}^{N-4}\psi_j-i\psi_{N-1} \psi_N-i\psi_{N-3} \psi_{N-2}\\
&-\psi_{N-3} \psi_{N-2}\psi_{N-1} \psi_{N}+(-i)^{(N-2)/2}\psi_{N-1} \psi_{N}\prod_{j=1}^{N-4}\psi_j)
\end{align*}
\normalsize
After rewriting the perturbation term using Majorana fermions, the joint moments of $H_{\text{SYK}}$ and $D_c$ can be calculated through commutative relations between the Majorana fermions. As $D_c$ is a constant term, the ensemble averaging only affects $H_{\text{SYK}}$, meaning that pairing only occurs within the index sets of $H_{\text{SYK}}$. We typically assign a weight $q$, defined in \eqref{q} to the intersection between $H_{\text{SYK}}$ as usual. However, it becomes necessary to compute the intersection between $H_{\text{SYK}}$ and $D_c$ as well, since $D_c$ is written in terms of Majorana fermions, and the corresponding average factor is denoted by $\tilde{q}(r)$, depending on $r$, the fraction of non-zero entries in $D_c$. For example, when $c=2^{(N-4)/2}$, 
\begin{align*}
\tilde{q}(1/4)=\frac{1}{2}\left(1+\frac{1}{\binom{N}{p}}\sum_{m=0}^2(-1)^{m}\binom{2}{m}\binom{N-2}{p-m}\right)
\end{align*}
The results of the moment calculation presented here apply not only to the SYK model affected by a constant diagonal matrix, but also to any SYK model with a perturbation term that has a non-commutative crossing relation with the SYK model with an additional averaged weight factor $\tilde{q}$. The specific characteristics of the perturbation term define $\tilde{q}$, although it does not influence the moment calculation in the primary text. As a result, the calculation of $\tilde{q}$ for the diagonal matrix with constant value $\theta$ has been provided in Appendix \ref{appd:2}. Subsequently, we may regard $\tilde{q}$ as a fixed value.


This study aims to establish connections between the eigenvalue distribution of the perturbed SYK model and non-commutative probability theory. Our focus lies particularly on computing mixed moments involving the DSSYK model and the constant term. Motivation for exploring non-commutative probability theory stems from the established link between the DSSYK model and $q$-Brownian motion \cite{Speicher1}. The eigenvalue distribution of the SYK model exhibits a shift from a Gaussian distribution ($q=1$) to a Wigner semicircular law ($q=0$). This transition reflects a correlation between the central limit theorem properties observed for classical and free random variables, similar to the behavior of a $q$-Gaussian random variable. In a related context, \cite{Berkooz3} introduces the $q$-Gaussian element $T$ (detailed in section \ref{sec:2}) as an ``effective Hamiltonian" for the DSSYK model, operating on a Hilbert space of chord diagrams. This crucial characterization enables us to employ a non-commutative probability approach to SYK-like models. When a diagonal matrix with a single non-zero entry is introduced to the DSSYK model, the resulting distribution of eigenvalues shows a phase transition similar to that of a random matrix, see \cite{Wu}. This phenomenon can be explained by the free convolution theory between the spectral measure of the initial model and the spectral measure of the perturbation term.

Another motivation arises from recent literatures on the DSSYK model and its non-commutative gravity dual, as outlined in \cite{Lin, Berkooz3} and related references. These studies emphasise that chord diagrams are more than just mathematical tools for computing the SYK eigenvalue distribution, they are seen as the appropriate framework for describing the bulk theory of its gravity dual. Additionally, it should be pointed out that several gravity models are related to free probability theory (see \cite{Wang} for an example). Therefore, this study has the potential to provide valuable insights into these areas of research.
Finally, it is worth noting that while our results lie within the \eqref{double-scaled} limit, they also show a significant agreement with fixed finite $p$ cases (as observed in a comparable situation by \cite{Garcia1, Garcia2}). Section \ref{sec:6} will present a comparison of our analytical results with the numerical computation of the $p=4$ case.
\subsection{Results}
\label{subsec:1}
As in \cite{Wu}, we denote $m_n$ as the ensemble averaging moments of \eqref{hamiltonian} extracted by ensemble averaging moments of the SYK Hamiltonian
\begin{align}
m_n=\frac{1}{r}\left(\langle \text{tr}H^n\rangle_J-\langle \text{tr}H_{\text{SYK}}^{2\lfloor n/2\rfloor}\rangle_J\right)
\label{rm}
\end{align}
Recall that parameter $r$ is defined as the proportion of non-zero entries on the diagonal of $D_c$. We divided by $r$ to eliminate $r$ from the resulting moment formula; instead, the dependence on $r$ is encoded in $\tilde{q}$. The main results of the paper are
\small
\begin{align}
m_n&=\nonumber\nonumber\\
&\sum_{j=0}^{\lfloor\frac{n-1}{2}\rfloor}\frac{n\theta^{n-2j}}{n-2j}\sum_{\substack{l=1 \\ k_1+k_2+\cdots + k_l = 2j\\k_i\neq 0 \forall i\in[1,l]}}^{\min[p-2j,2j]}\binom{n-2j}{l}\prod_{i=1}^l\sum_{m_i=0}^{\lfloor k_i/2\rfloor} c_{m_i,k_i}\sum_{n_{xy}}\prod_{x=1}^{l}\binom{k_i-2m_i}{n_{x1},\dots,n_{xl}}_q\prod_{1\leq k<p\leq l}[n_{kp}]_q!\tilde{q}^{n_{kp}}q^{B}
\label{mm}
\end{align}
\normalsize
where $[n]_q!=\prod_{l=1}^n [k]_q$ denotes the $n$ $q$-factorial with $[k]_q=(1-q^k)/(1-q)$.
The sum $\sum  n_{xy}$  is  over all non negative symmetric matrices $n_{xy}$ with integer entries satisfying $n_{xx}=0$, $\sum_{x=1}^ln_{xy}=k_y-2m_y$ with $y=1,2,\dots,l$ and $B=\sum_{1\leq i<j<m<p\leq  l}n_{im}n_{jp}$. This sum actually corresponds to the linearization coefficients of the product of $q$-Hermite polynomials:   
\begin{equation}
\int_{-\infty}^{\infty}  \prod_{j=1}^{k} H^{(q)}_{n_{j}}(x)\nu_q(x)dx=\sum_{n_{i j}} \prod_{i=1}^{k}\left(\begin{array}{c}n_{i} \\ n_{i 1}, \ldots, n_{i k}\end{array}\right)_{q} \prod_{1 \leq i<j \leq k}[n_{i j}]_q! q^{B}
\label{coeffH}
\end{equation}
where the $q$-Hermite polynomial $H^{(q)}_n(x)$ is defined in equation \eqref{q-H} and its orthogonal measure $\nu_q(x)$ is defined in equation \eqref{densitySYK}.
Equation \eqref{coeffH} counts the  number of perfect matchings in the complete graph $k_n$ with $n$ vertices divided into $k$ subsets, where $n=n_1+n_2+\dots+n_k$  and all the matchings are inhomogeneous (for each edge, the two vertices come from different subsets), factor q enumerates the number of crossing between edges \cite{Viennot}.
The coefficients 
\begin{align}
c_{m, n}:=\frac{1}{(1-q)^m}\sum_{j=0}^{m}(-1)^{j} q^{j+\binom{j}{2}} \frac{n-2 m+2 j+1}{n+1}\left(\begin{array}{c}n+1 \\ m-j\end{array}\right)\left(\begin{array}{c}n-2 m+j \\ j\end{array}\right)_{q}
\label{opc}
\end{align}
enumerates the number of chord diagrams with $n$ nodes, $m$ chords and $n-2m$ open chords (chords whose ends are not yet be connected by another node) according to the number of crossings between chord-chord and  chord-open chord.
The moment formula \eqref{mm} can be obtained from its generating function:
\begin{align}
m_n&=[z^n]\langle 0\vert z\frac{d}{dz} \log\frac{1}{1-B(z,x_0)}\vert 0\rangle\nonumber\\
&=n[z^n]\langle 0\vert  \log\frac{1}{1-B(z,x_0)}\vert 0\rangle
\label{gf1}
\end{align}
here, $\langle 0\vert .\vert 0\rangle$ denotes the vacuum expectation with respect to the $q$-Gaussian distribution $\nu_q(x_0)$. The function $B(z,x_0)$ is defined as
\small
\begin{align}
B(z,x_0)&=\nonumber\\
&\theta\sqrt{1-q}\sum_{k\geq 0}\frac{\sqrt{\tilde{q}(1-q)}^k H_{k}^{(q)}(x_0)}{[k]_q!}\sum_{j\geq0}(-1)^j\frac{[k+j]_q!}{[j]_q!}q^{j(j+1)/2}\left( \frac{1-\sqrt{1-4z^2/(1-q)}}{2z/\sqrt{1-q}}\right)^{2j+k+1}
\end{align}
\normalsize
We will see in Section \ref{sec:4} that this function is the generating function of a conditional expectation corresponding to the $q$-Ornstein-Uhlenbeck process.
\subsection{Plan of the paper}
\label{subsec:2}
Section \ref{sec:2} is a standalone review of established associations between the original DSSYK model and $q$-Gaussian random variables, which provides insight into the suitability of $q$-Gaussian variables as a DSSYK approximation. Additionally, these results are pertinent to our moment calculations. In section \ref{sec:3}, we revisit results of \cite{Wu}, which in the context of this paper is a special case $c=1$, and we establish links with non-commutative probability theory. Section \ref{sec:4} provides a thorough explanation of calculating moment with arbitrary $c$. This can be achieved by solving the enumeration problem involving the same combinatorial structure identified in \cite{Wu}, but with different building blocks. The change of the building block is identified with a $q$-Ornstein-Uhlenbeck ($q$-OU) process.
In section \ref{sec:5}, we introduce a joint moment-cumulant relation that recovers our moment calculation results. Using this relation, we explain the implications of adding a constant term $D_c$ in the context of non-commutative probability theory. Section \ref{sec:6} presents some numerical results, we compare our analytic expressions with the numerical results obtained by calculating the eigenvalues of \eqref{hamiltonian} with Mathematica. The numerical evidence reveals a phase transition in the eigenvalue spectrum depending on the values of $r$ and $c$. Finally, section \ref{sec:7} provides a brief conclusion and outlines possible future research directions.

\section{Double-scaling limit SYK model and $q$-Gaussian Processes}
\label{sec:2}
In this section, we review some of the results in \cite{Berkooz1,Berkooz2} and \cite{Speicher1} which connect the DSSYK model to the $q$-Gaussian Processes. We further establish a concrete connection between the $T$ operator in \cite{Berkooz1,Berkooz2} and the $q$-Hermite polynomials. 

\begin{figure}[H]
\begin{center}
\includegraphics[scale=.7]{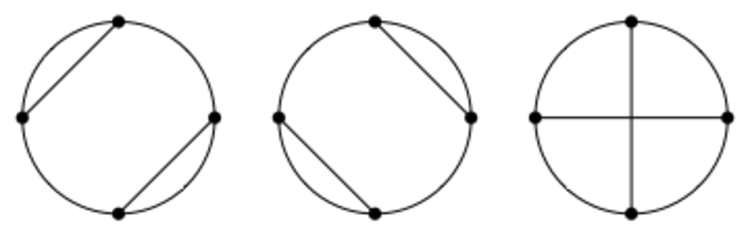}
\caption{$3$ possible chord diagrams of size $4$. $\langle\text{tr} H_{\text{SYK}}^4\rangle_J=2+q$}
\label{cd4}
\end{center}
\end{figure}

\begin{figure}[H]
\begin{center}
\includegraphics[scale=0.9]{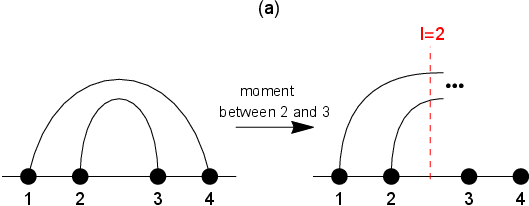}
\includegraphics[scale=0.9]{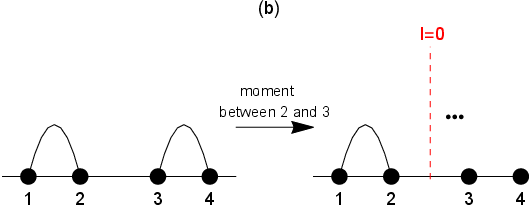}
\caption{An example for open chords between two adjacent nodes}
\label{cutopen}
\end{center}
\end{figure}

The moments calculation of DSSYK in \cite{Berkooz1,Berkooz2} is mapped into a perfect matching enumeration problem in chord diagrams,
$$
\langle \text{tr} H_{\text{SYK}}^{2k}\rangle_{J}=\sum_{\pi\in\mathcal{P}_2(2k)}q^{cr(\pi)}
$$
As an example, figure \ref{cd4} shows all possible chord diagrams of four nodes enumerated by $\langle\text{tr} H_{\text{SYK}}^4\rangle_J$.
It is also possible to further cut the circle open and place the chord diagram on the upper half plane, as illustrated in  figure \ref{cutopen}. 
Considering an evolution from the left to the right of a chord diagram, each node in the diagram represents a $H_{\text{SYK}}$, a chord connect two $H_{\text{SYK}}$ who share the same index set. At any given moment (an interval between two successive nodes), one can define a state $\vert l\rangle$ with $l=0,1,2,...$ as representing the number of open chords (chords that are not yet closed) at that moment. Figure \ref{cutopen} shows an example for open chords. It is evident that there are no open chords before the first node and after the last node. 
The left-hand sides of $(a)$ and $(b)$ in figure \ref{cutopen} represent the two possible non-crossing parings in $\langle\text{tr} H_{\text{SYK}}^4\rangle_J$. If we consider the interval between the second and third nodes, the number of open chords, $l$, can be either two or zero.
Each time we insert a node, it can either open a new chord or close a old chord. If it's the later case, the closed chord can be intersected with the previous open chords with each intersection gives a factor of $q$.
Such  evolution can be captured by a transfer matrix acts on the Hilbert space of chords
$$
T\vert l\rangle =\vert l+1\rangle+[l]_q\vert l-1\rangle.
$$
Thus
$$
\langle \text{tr} H_{\text{SYK}}^{2k}\rangle_{J}=\langle 0\vert T^{2k} \vert 0\rangle
$$
since the start and the end of a chord diagram have zero chords. 
This $T$ transfer matrix turns out to be a particular realization of a $q$-Gaussian element :
\begin{align}
T=a+a^{\ast}
\label{q-g}
\end{align}
where $a^{\ast}$ and $a$ are creation and annihilation operators satisfying the $q$-commutative relation
\begin{align}
aa^{\ast}-qa^{\ast}a=\mathbb{1}
\label{car}
\end{align} 
where $\mathbb{1}$ denotes the identity operator. These operators acting on the Hilbert space $\vert l\rangle$ with a vacuum $\Omega$ such that
\begin{align*}
\Omega &=\vert 0\rangle\\
a\vert 0\rangle &= 0\\
a^{\ast}\vert 0\rangle &=\vert 1\rangle\\
a\vert n\rangle &= [n]_q \vert n-1\rangle\\
a^{\ast}\vert n\rangle &= \vert n+1\rangle\\
a^{\ast}a\vert n\rangle &=[n]_q\vert n\rangle.
\end{align*}
Moreover, one can define a number operator $\hat{N}$ such that
\begin{align}
\hat{N}\vert n\rangle=n\vert n\rangle\text{ with }[a,a^{\ast}]=q^{\hat{N}}
\label{nop}
\end{align}

\begin{figure}[H]
\begin{center}
\includegraphics[scale=0.4]{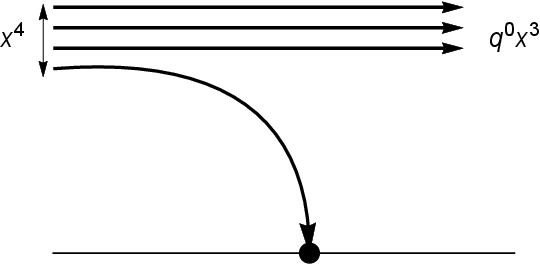}
\includegraphics[scale=0.4]{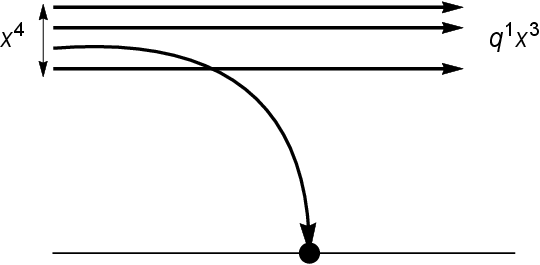}
\includegraphics[scale=0.4]{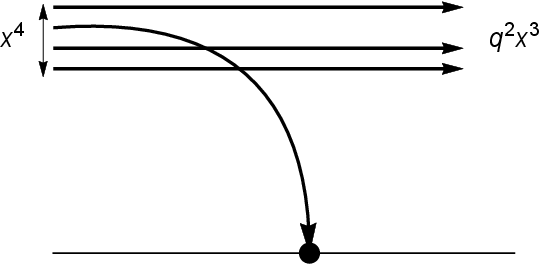}
\includegraphics[scale=0.4]{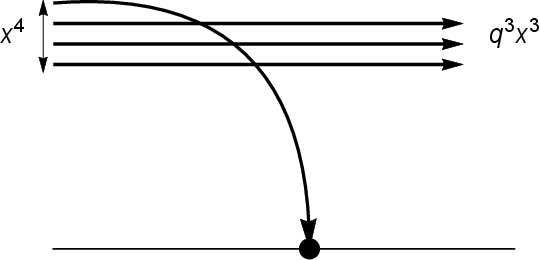}
\caption{Graphic illustration of $\bigtriangleup_q x^4$}
\label{dx4}
\end{center}
\end{figure}

If we have $n$ copies of $T$ acting on the Hilbert space $\vert l\rangle$ (which is a vector space of chords), each of them can either open ($a^{\ast}$) a new chord or close ($a$) an old chord, when it closes an old chord, we pair it to the previous $T$ who opened this chord. Notice that we can choose a specific realization of the creation-annihilation  axiom \eqref{car}: the creation and annihilation operators $a^{\ast}$ and $a$ can be understood as multiplication and $q$-difference operators respectively acting on the chords space
\begin{align*}
&a^{\ast}\rightarrow x\\
&a\rightarrow \bigtriangleup_q\quad\text{where}\quad \bigtriangleup_q f(x)=\frac{f(qx)-f(x)}{(q-1)x}\quad\text{with}\quad q\in[0,1].
\end{align*}
Since we are in chord space, $x$ represents an elementary component (``atom'') of chord diagrams, an open chord. 
Let $f$ be a smooth function of $x$, which can be thought of as a 
as the generating function of an arbitrary class of combinatorial objects whose elements consist of chords (x-atoms). The $q$-difference operator becomes the standard derivative operator when $q\rightarrow 1$, and it acts linearly on the monomial $x^n$ 
\begin{align}
\bigtriangleup_q x^n=[n]_q x^{n-1}\quad\text{with}\quad [n]_q:=1+q+\dots+q^{n-1}=\frac{1-q^n}{1-q}
\label{q-diff}
\end{align}
Combinatorially, $x^n$ can be regarded as $n$ open chords, $
\bigtriangleup_q$ acting on it means we pick one of the chords to close, and during the process, the closed chord can be interacted by the other $n-1$ open chords, we record each crossing by a $q$, therefore the exponent of $q$ equals the number of crossing between this closed chord and the other open chords. For example, when $n=4$ (see figure \ref{dx4})
\begin{align*}
\bigtriangleup_q x^4=\overbrace{\cancel{x} xxx}^{q^0} +\overbrace{x\cancel{x}xx}^{q^1}+\overbrace{xx\cancel{x}x}^{q^2}+\overbrace{xxx\cancel{x}}^{q^3}=[4]_q x^3
\end{align*}
We refer \cite{Blasiak} for more detailed introduction about the relation between creation annihilation axiom and the representation of the Weyl relations.
Thus, replacing $a^{\ast}$ and $a$ by $x$ and $\bigtriangleup_q$ respectively, we can expand the operator $T^n$ and reduce it to a unique normal form by using the rule \eqref{q-diff} where in each term, all the occurrences of $x$ precede all the occurrences of $\bigtriangleup_q$. For example 
\begin{align*}
T^2=(\bigtriangleup_q+x)^2&=\bigtriangleup_q^2+
\bigtriangleup_qx+x\bigtriangleup_q+x^2\\&=x^2+(1+q)x\bigtriangleup_q+\bigtriangleup_q^2+1\\
&=\normord{T^2}+1
\end{align*}
where $\normord{T^2}$ is denoted as the normal-ordering of $T^2$.  For higher powers of $n$, we have
\begin{align*}
T^3&=\normord{T^3}+(2+q)\normord{T}\\
T^4&=\normord{T^4}+(3 + 2 q + q^2)\normord{T^2}+(2+q)
\end{align*}
it turns out that the normal ordering   $\normord{T^n}=H_{n}^{(q)}(T)$ satisfying the recurrence relation (see full proof in \cite{Bozejko})
\begin{align}
xH_{n}^{(q)}(x)=H_{n+1}^{(q)}(x)+[n]_qH_{n-1}^{(q)}(x)
\label{q-H}
\end{align}
where $H_{n}^{(q)}(x)$ are the $q$-Hermite polynomials with initial conditions $H_{0}^{(q)}(x)=1$ and $H_{1}^{(q)}(x)=x$. $H_{n}^{(q)}(x)$ are orthogonal with respect to the $q$-Gaussian distribution\footnote{Note that $H_{n}^{(q)}(x)$ here is a modified version of the continuous $q$-Hermite polynomial $H_{n}(x\vert q)$, satisfying $$H_{n}^{(q)}(x)=H_{n}(x\sqrt{1-q}/2 \vert q)/(1-q)^{n/2} .$$ Here, it is preferable to utilise $H_{n}^{(q)}(x)$ as it carries a combinatorial meaning. Further details can be found in \cite{Viennot}. One can conjugate the transfer matrix $T$ to a symmetric version, then the corresponding orthogonal polynomial will be the continuous $q$-Hermite polynomial $H_{n}(x\vert q)$, see section $3.2$ in \cite{Berkooz1}} 
\small
\begin{align}
\nu_{q}(dx)=\frac{\sqrt{1-q}}{4\pi\sin\theta} \prod_{k=1}^{\infty}(1-q^k)\vert 1-q^k e^{2i\theta}\vert^2\quad\text{with}\quad x=\frac{2\cos\theta}{\sqrt{1-q}}\quad\forall\theta\in[0,\pi]
\label{densitySYK}
\end{align}
\normalsize
Therefore,  
\begin{align}
T^k=\sum_{m=0}^{\lfloor \frac{k}{2}\rfloor}c_{m,k}H_{k-2m}^{(q)}(T)
\label{top}
\end{align}
the coefficient $c_{m,k}$ is defined in \eqref{opc}, it comes from the contractions between $\bigtriangleup_q$ and $x$ (the ``$1$" in $\bigtriangleup_q x\rightarrow 1+qx\bigtriangleup_q$). These contractions reduce the degree of the monomials, $c_{m,k}$ is the coefficient of a reduced term of degree $k-2m$, it depends on $q$ according to the rule \eqref{q-diff}. We need to sum over all possible sets of contractions,
 in general, we have (see \cite{ANSHELEVICH1}, Proposition 6.12, Construction 3.7 and Section 6.2)
\begin{align}
T^k=\sum_{\pi\in\mathcal{P}_{1,2}(k)} q^{cr(\pi)+sd(\pi)}H_{s_1(\pi)}^{(q)}(T)
\label{qGp}
\end{align} 
where the sum is over all the partition $\mathcal{P}_{1,2}(k)$ of set $\lbrace 1,2,\dots k\rbrace$ with each block has at most $2$ elements (we have a block with $2$ elements whenever there is a contraction between $\bigtriangleup_q$ and $x$, otherwise the blocks are singletons). This operator depends on the number of crossing $cr(\pi)=\vert \lbrace i,j\in [k]: i<j<\pi(i)<\pi(j)\vert\rbrace+\vert \lbrace i,j\in [k]: \pi(j)<\pi(i)<j<i\vert\rbrace$, $sd(\pi)=\sum_{i=1}^{s_1(\pi)}d(i)$ with the depth of the $i$-th singleton  $d(i)=\vert \lbrace j\vert \exists a,b\in B_j, a<i<b\rbrace\vert$ where $B_j$ are the $2$-elements blocks in $\pi$  and $s_1(\pi)$ is the number of singleton in $\pi$ \footnote{For example, $\pi =\lbrace (1,3),(2,5),(4)\rbrace \in\mathcal{P}_{1,2}(5)$, $cr(\pi)=1$, $s_1(\pi)=1$, there is only one singleton $(4)$, therefore $d(4)=1$ and $sd(\pi)=d(4)=1$.}. Therefore the vacuum expectation value of $T^n$ is
\begin{align}
\langle 0\vert T^k\vert 0\rangle=\sum_{\pi\in\mathcal{P}_2(k)}q^{cr(\pi)}
\label{veT}
\end{align}
since the normal-ordered part is precisely the part that will vanish when one takes the vacuum expectation value. In the context of non-commutative probability theory, the operators act on some Hilbert spaces having a meaning as non-commutative analogues of the probabilistic notions of random variables. The vacuum expectation corresponds to the expectation of the corresponding random variable with respect to its distribution $\langle 0\vert \cdot \vert 0\rangle =\mathbb{E}[\cdot]$. Therefore, \eqref{veT} is equivalent to
\begin{align}
\int x^k \nu_q({dx})
\end{align}
where $\nu_q(dx)$ is the $q$-Gaussian distribution defined in \eqref{densitySYK}. This requires that all partitions are perfect matchings of $[k]$ with an even $k$, otherwise the result becomes zero. Remarkably, this is consistent with the ensemble averaging $k$-th moment of the DSSYK model. This observation leads to the consideration that after the ensemble average, the DSSYK model converges to a $q$-Gaussian random variable in the sense of its eigenvalue distribution.
\section{Moments calculation: $c=1$}
\label{sec:3}
In this section, we undertake a comprehensive re-examination of the previous results concerning the only one nonzero entry perturbation term. Subsequently, we reinterpret the moment formula to establish a connection with non-commutative probability theory.

In \cite{Wu}, by mapping the ensemble averaging moments calculation of \eqref{hamiltonian} into a counting problem of some complex combinatorial structure, we find the eigenvalue distribution of our perturbative SYK model, whose moments are given by
\begin{align}
m_n =\sum_{j=0}^{\lfloor\frac{n-1}{2}\rfloor}\theta^{n-2j}\frac{n}{n-2j}\sum_{\substack{k_1,k_2,\ldots,k_j \geq 0 \\ k_1+2k_2+\cdots + jk_j = j}} \binom{n-2j}{k_1,k_2,\ldots,k_j,n-2j-\sum_{l=1}^j k_l} \prod_{i=1}^j \text{RT}(i,q)^{k_i}
\label{moment2}
\end{align}
where $\text{RT}(i,q)=\text{tr}\langle H_{\text{SYK}}^{2i}\rangle_J$ is the Riordan-Touchard formula \cite{Riordan, Touchard} that computes the ensemble-averaged $2i$-th moment of SYK in the double-scaling limit \cite{Berkooz1,Garcia2,Erdos}.
We can rewrite this as
\small
\begin{align}
m_n &=\nonumber\\
&\sum_{j=0}^{\lfloor\frac{n-1}{2}\rfloor}\frac{n}{n-2j}\sum_{\substack{k_1,k_2,\ldots,k_j \geq 0 \\ k_1+2k_2+\cdots + jk_j = j}} \binom{n-2j}{k_1,k_2,\ldots,k_j,n-2j-\sum_{l=1}^j k_l} \theta^{n-2j-\sum_{l}k_l}\prod_{i=1}^j (\theta\text{RT}(i,q))^{k_i}
\label{moment3}
\end{align}
\normalsize
Note that the second sum resembles the Boolean moments-cumulants relation. The notion of the Boolean cumulant and its underlying combinatorial aspect was first introduced in \cite{Speicher5}. Namely, if we define the generating functions of moments and Boolean cumulants as $M(z)=\sum_{k\geq0} m_k z^k$ and $H(z)=\sum_{k\geq0} h_k z^k$ respectively, the Boolean moment-cumulants relation is given by:
\begin{align}
m_n=\sum_{k_1+2k_2+\dots+n k_n=n}\binom{k_1+k_2+\dots+k_n}{k_1,k_2,\dots,k_n}\prod_{i=1}^n h_i^{k_i}
\label{moments-cumulants}
\end{align}
this relation comes from the fact that the Boolean independent random variables $a_i$ with $i=1,2,\dots ,n$ satisfying 
\begin{align}
\varphi( a_1 a_2\dots a_n)=\varphi(a_1)\varphi(a_2)\cdots\varphi(a_n)
\label{boolean}
\end{align}
where $\varphi(a_i^{k})$ being the $k$-th moment of $a_i$. The corresponding combinatorial description behind this is called an interval partition (\i.e. a partition with blocks that are intervals of consecutive numbers). To illustrate, consider the case with $3$ boolean independent variables $a_1$, $a_2$ and $a_3$. An example of their mixed moment can be expressed as follows:
$$
\varphi (a_1a_1a_2a_3a_3a_3a_2)=\varphi (a_1^2)\varphi (a_2)\varphi (a_3^3)\varphi(a_2)=\varphi (a_1^2)\varphi (a_2)^2\varphi (a_3^3)
$$
Obviously this is different from classical independence, the mixed moments here are factorised into the product of the individual moments of $a_1$, $a_2$ and $a_3$ along the intervals.

The alignment of our moment formula for the case $c=1$ with this relation results from the introduction of a constant perturbation to the SYK Hamiltonian, which creates a new form of non-commutative random variables. The number of $\theta$ in equation \eqref{moment3} limits the size of the interval partition. In this case we only count products of consecutive $H_\text{SYK}$, because the extra source $D_{1}$ has only one nonzero entry, the product of $D_{1}$ sandwiched between any number of paired $H_\text{SYK}$ can be neglected, therefore 
\begin{align}
m_n\sim\sum\text{RT}(1,q)^{k_1}\text{RT}(2,q)^{k_2}\cdots\text{RT}(l,q)^{k_l}\cdots \theta^{n-2\sum_{i}ik_i}
\label{nf1}
\end{align}
where the sum is over all combinations of products of $\text{RT}(i, q)$ and $\theta$ of different orders that satisfy the total exponent equal to $n$. 
Therefore, we recreate the relation \eqref{boolean} since $\text{RT}(k,q)=\langle \text{tr}H_{\text{SYK}}^{2k}\rangle_J$. See section 3 in \cite{Wu} for a specific calculation.
To conclude, by adding $D_1$, we distribute all $H_{\text{SYK}}$ in the mixed moments into intervals, and we use $\theta$ to mark each interval. We create independent relations between the intervals. Thus, each product of $H_{\text{SYK}}$ in any interval can be approximated by a Boolean random variable $a_i$. For example, if we have $2k$ $H_{\text{SYK}}$ distributed in the $i$-th interval, we have
$\theta\langle \text{tr}H_{\text{SYK}}^{2k}\rangle_J=\varphi(a_i)$.
\section{Moments calculation: general c}
\label{sec:4}
The aim here is to compute the reduced moments defined in \eqref{rm}, it is expanded as
\small
\begin{align}
m_n:&=\frac{1}{r}\sum_{j=0}^{\lfloor\frac{n-1}{2}\rfloor}\binom{n}{2j}\theta^{n-2j}\sum_{\alpha_1,\dots,\alpha_{2j}}\langle J_{\alpha_1}\dots J_{\alpha_{2j}}\rangle_J \text{tr}\Psi_{\alpha_1}\dots D_c\dots\Psi_{\alpha_k}\dots D_c\dots\Psi_{\alpha_{2j}} \nonumber\\
&=\frac{1}{r}\sum_{j=0}^{\lfloor\frac{n-1}{2}\rfloor}\binom{n}{2j}\theta^{n-2j}\sum_{\substack{l=1 \\ k_1+k_2+\cdots + k_l =2 j}}^{\min[n-2j,2j]} \text{tr}\Psi_{\alpha_1}\dots\Psi_{\alpha_{k_1}}D_c\dots D_c\Psi_{\alpha_{\sum_{i=1}^{l-1}k_i+1}}\dots\Psi_{\alpha_{\sum_{i=1}^lk_i}}D_c
\label{moments1}
\end{align}
\normalsize
Recall that $r$ is the ratio of non-zero entries in the diagonal of $D_c$, it comes from the normalised trace operator $\text{tr}$. 
The summation $\sum_{\alpha_1,\dots,\alpha_{2j}}$ covers all possible index sets. Ensemble averaging $\langle \dots\rangle_{J}$ pairs up the $\Psi_{\alpha}$ terms, leading to $j$ sets of indices post-averaging. These sets are counted by $\binom{N}{p}^j$ and compensated by $\langle J_{\alpha_1}\dots J_{\alpha_{2j}}\rangle_J$. With a total of $\sum_{i=1}^lk_i=2j$ $\Psi_{\alpha}$ elements, the total number of $D_c$ becomes $n-2j$ with $j\in [0,\lfloor n/2\rfloor]$. By introducing the constant terms $D_c$, we organize the $2j$ $\Psi_{\alpha}$ into $l$ intervals, where the maximum number of intervals is determined by $\min[n-2j,2j]$. If we view each interval as a building block, we can break down the calculation in \eqref{moments1} into two stages. Firstly, we must count the number of interval partitions that can be formed using these building blocks. Secondly, we need to determine the contribution of each building block. The partition of the intervals shown in equation \eqref{moments1} has previously been identified as a pointed cyclic combinatorial structure in section 2 of \cite{Wu}.  The structure exhibits cyclic invariance during rotation due to the trace operator, and its enumeration can be done as follows.

\begin{figure}[H]
\begin{center}
\includegraphics[scale=1.]{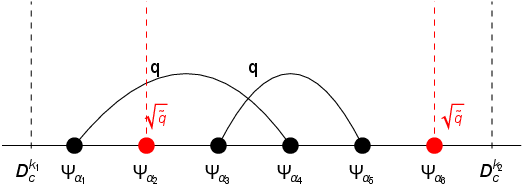}
\caption{An interval with a partition $\lbrace (1,4),(2),(3,5),(6)\rbrace$ inside: two pairs $\lbrace(\Psi_{\alpha_2},\Psi_{\alpha_4}),(\Psi_{\alpha_3},\Psi_{\alpha_4})\rbrace$ along with two singletons, $\Psi_{\alpha_4}$ and $\Psi_{\alpha_6}$. In this diagram, there are two types of crossings, both marked by $q$: one occurs between two chords, and the other takes place between a chord and an open chord. The overall contribution from this interval is $\tilde{q}q^2$.}
\label{int_p}
\end{center}
\end{figure}

\begin{figure}[H]
\begin{center}
\includegraphics[scale=1.]{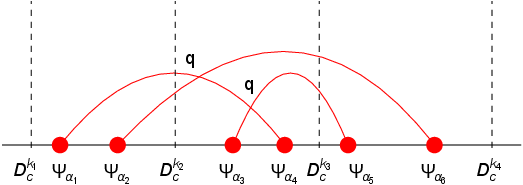}
\caption{An example of parings between three intervals: it contributes $q^2$}
\label{int_bw}
\end{center}
\end{figure}

Consider a complex combinatorial structure denoted $\mathcal{A}$, which comprises a pointing operator applied to labelled cycles within a rooted substructure $\mathcal{B}$.  In other words, within $\mathcal{A}$, each element is a pointed necklace made up of labelled nodes, and each node is connected to elementary building blocks that belong to $\mathcal{B}$. The corresponding generating function for $\mathcal{A}$ is
\begin{align}
\mathcal{A}=\Theta\operatorname{CYC}\left(\mathcal{Z}\times\mathcal{B}\right)\quad\Longrightarrow \quad A(z)=z\frac{d}{dz}\log\frac{1}{1-zB(z)}
\label{op_complex}
\end{align}
where $\Theta$ denotes the pointing operator, $\operatorname{CYC}$ corresponds to a labelled cycle structure and $\mathcal{Z}$ symbolizes the root within in $\mathcal{A}$. $A(z)=\sum_{k}a_k z^k$  and $B(z)=\sum_{k}b_k z^k$ are two generation functions enumerate structures $\mathcal{A}$ and $\mathcal{B}$ respectively. Each operator in a combinatorial structure can be directly translated into operations applied to its corresponding generating function. For comprehensive definitions of these operators in combinatorial structures and their corresponding enumerations via generating functions, see \cite{Flajolet2}. To come back to our case, the substructure $\mathcal{B}$ here corresponds to the intervals containing $\Psi_{\alpha}$. To calculate $A(z)$, we must determine the contribution of each interval. Consider the $i$-interval, denoted as $(\Psi_{\alpha_{k_1+\dots+k_{i-1}+1}},\dots,\Psi_{\alpha_{k_1+\dots+k_i}})$, there are $k_i$ $\Psi_{\alpha}$. These $\Psi_{\alpha}$ can either be paired if they share the same index set or remain unpaired if their index sets are unique within the interval. Thus, in the $i$-th interval, we have a partition $\pi_i \in \mathcal{P}_{1,2}(k_i)$, where $\mathcal{P}_{1,2}(k_i)$ represents a partition of $[k_i]=\lbrace 1,2,\dots,k_i\rbrace$, with each block containing at most $2$ elements. The remaining unpaired $\Psi_{\alpha}$ are called to as singletons and will be paired with singletons from other intervals, since all $\Psi_{\alpha}$ must be paired due to the ensemble averaging $\langle \dots\rangle_{J}$. These pairings between singletons from different intervals are called ``inhomogeneous" with respect to the interval partition \cite{Leroux}. 
We propose a new chord diagram description, where each $\Psi_{\alpha}$ corresponds to a node, while a pair of $\Psi_{\alpha}$ is represented as a chord connecting these nodes. The presence of one or more $D_c$ between two nodes is shown as a ``wall" and is represented by a dotted line in the diagram. We use $q$ to mark a crossing between chords and $\sqrt{\tilde{q}}$ to mark a singleton, so each chord crossing a ``wall" contributes a $\tilde{q}$. Note that a pair of $\Psi_{\alpha}$ crossing through a $D_c$ or a number of successive $D_c$ will both contribute only one $\tilde{q}$ because $D_c$ is constant. See the figures \ref{int_p} and \ref{int_bw} as examples to illustrate these two types of pairing. We distinguish elements in the diagram as follows: red points represent singletons, black points denote paired nodes, black curves represent chords closed inside intervals, red dotted lines signify open chords inside each interval, black dotted lines represent ``wall"s, and red curve connecting singletons crossing the ``wall"s. It should be emphasized that the contribution of the crossing between the chords and the ``wall"s (the power of $\tilde{q}$) is enumerated within each interval, as we have already marked each singleton with a $\sqrt{\tilde{q}}$.

To summarize, after partitioning \(\Psi_{\alpha}\) into intervals, we perform ensemble averaging pairings in two steps: first within each interval and then for the remaining singletons between different intervals. This process divides the calculation in \eqref{moments1} into two stages:
\begin{itemize}
\item Within each interval, pairing follows the same pattern as in the original DSSYK model, which can be enumerated by a transfer matrix $T$ in the chord Hilbert space (see section \ref{sec:2}). We use $\sqrt{\tilde{q}}$ to mark each occurrence of the singleton. This is because all singletons are subsequently paired, and each pairing between singletons adds an extra weight of $\tilde{q}$ due to the passage through $D_c$. The number of singletons can be determined as $k_i-2m_i$, where $m_i$ represents the number of 2-element blocks (pairs) in $\pi_i$. Thus for  the $i$-th interval, we have
\begin{align}
\sum_{\pi_{i}\in \mathcal{P}_{1,2}(k_i)}q^{cr(\pi_i)+sd(\pi_i)}\tilde{q}^{(k_i-2m_i)/2}H_{k_i-2m_i}^{(q)}(T)
\label{stoch}
\end{align}
$H_{k_i-2m_i}^{(q)}(T)$ is the $k_i-2m_i$-th $q$-Hermite polynomial of $T$. When $\tilde{q}=1$, this formula coincides with \eqref{qGp}, since the pairing mechanism within each interval mirrors that of the DSSYK model. This step is actually the building block of our interval partition, leading to the fact that $B(z)$ in \eqref{op_complex} is given by \eqref{stoch}.
In the following section, we'll explain how the transition from \eqref{qGp} to \eqref{stoch} can be understood as a $q$-Ornstein-Uhlenbeck ($q$-OU) process, with \(B(z)\) representing the generating function of conditional expectation moments.
It's worth noting that \eqref{stoch} is a sum of partition-dependent stochastic measures of the $q$-Brownian motion as defined in \cite{ANSHELEVICH1,ANSHELEVICH2}.
\item Between the interval, we need another partition $\tilde{\pi}_i \in \mathcal{P}_{2}(\sum_{i=1}^l k_i-2m_i)$ to connect all those $l$ intervals. $\tilde{\pi}_i$ is an inhomogeneous pairing with respect to interval partition $\pi_{k_1-2m_1,k_2-2m_2,\dots,k_l-2m_l}$
\footnote{An interval partition $\pi_{n_1,n_2,\dots,n_k}$ is a set partition of $\lbrace 1,2,\dots,2n\rbrace$ whose blocks are intervals of consecutive integers of lengths
$n_1, n_2,...,n_k$ with $\sum_{j=1}^k n_k=2n$. An inhomogeneous pairing with respect to $\pi_{n_1,n_2,\dots,n_k}$ is a set of pairs $\lbrace (a(i),b(i))\rbrace_{i=1}^{n}$ where $a(i)\in [n_{i_1}]$ and $b(i)\in [n_{i_2}]$ with $n_{i_1}\neq n_{i_2}$ and $n_{i_1}, n_{i_2}\in \lbrace n_1,n_2,\dots,n_k\rbrace$ .}. This is enumerated by
\begin{align}
\mathbb{E}\left[\prod_{i=1}^lH_{k_i-2m_i}^{(q)}(x)\right]=\sum_{\tilde{\pi}_i\in\mathcal{P}_{2}(\sum_{i=1}^l k_i-2m_i)}q^{cr(\tilde{\pi}_i)}
\label{expH}
\end{align}
$\mathbb{E}[\cdot]$ means the expectation value with respect to the measure of $q$-Hermite polynomial $H_{n}^{(q)}(x)$. It is worth emphasizing that the crossing between singletons is still weighted by $q$ because we mark any chord intersection with it. Additionally, the interaction between chords and ``wall"s is already accounted for by marking singletons with $\sqrt{\tilde{q}}$ beforehand. Since this step is post interval paring, it becomes necessary to compute an additional expectation value in the context of generating function enumeration in \eqref{op_complex} with respect to the $q$-Gaussian measure, $\langle 0\vert A(z)\vert 0\rangle$ due to the presence of coefficients in $B(z)$ that involve $q$-Gaussian variables.
\end{itemize}

As can be seen from the discussion above, the introduction of $D_c$ serves to effectively divide the ensemble averaging $\langle\cdots\rangle_{J}$ of the moment calculation in \eqref{moments1} into two distinct steps. 
This division gives rise to a contraction from $T^{k_i}$ to \eqref{stoch} within each interval containing $k_i$ nodes. In particular, this transformation corresponds to a map:
\begin{align}
H_{n}^{(q)}(x)\rightarrow\tilde{q}^{n/2}H_{n}^{(q)}(x)
\label{transf}
\end{align}
It transforms the $q$-Hermite polynomial into its multiplier. This can be understood as the existence of two types of open chords: those within each interval, denoted by $f_1$, and those outside the interval, denoted by $f_0$. The new chord Hilbert space is thus constructed using basis vectors $\vert f_0\rangle$ and $\vert f_1\rangle$ such that
\begin{align*}
&a_i\vert 0\rangle=0\\
&a^{\ast}_i\vert 0\rangle=\vert f_i\rangle\\
&a^{\ast}_i \left(\vert f_{i_1}\rangle\otimes\vert f_{i_2}\rangle\otimes\cdots\vert f_{i_n}\rangle\right)=\vert f_{i}\rangle\otimes\vert f_{i_1}\rangle\otimes\vert f_{i_2}\rangle\otimes\cdots\otimes\vert f_{i_n}\rangle\\
&a_i \left(\vert f_{i_1}\rangle\otimes\vert f_{i_2}\rangle\otimes\cdots\vert f_{i_n}\rangle\right)=\sum_{k=1}^n q^{k-1}\langle f_i\vert f_{i_k}\rangle \left(\vert f_{i_1}\rangle\otimes\vert f_{i_2}\rangle\otimes\cdots\otimes\vert \check{f}_{i_k}\rangle\otimes\cdots\otimes\vert f_{i_n}\rangle\right)
\end{align*}
with $i\in\lbrace 0,1\rbrace$ and $\vert \check{f}_{i_k}\rangle$ means $\vert f_{i_k}\rangle$ is deleted. The inner product is defined as
\begin{align}
\langle f_{i_j}\vert f_{i_k}\rangle=\sqrt{\tilde{q}}^{\vert i_j-i_k\vert}
\end{align}
The transformation \eqref{transf} corresponds to a $q$-OU process, denoted by $\mathbf{X}$. Considering two random variables, $x_{0},x_{1}\in\mathbf{X}$, both following $q$-Gaussian distribution $\nu_q$ and having a covariance is given by $\sqrt{\tilde{q}}$:
\begin{align}
&x_0=a_0+a^{\ast}_0;\text{ } x_1=a_1+a^{\ast}_1\quad\text{and }a_0a^{\ast}_1-qa^{\ast}_1a_0=\sqrt{\tilde{q}}\mathbb{1}\nonumber\\
&\langle 0\vert x_i^n\vert 0\rangle=\int x^n \nu_q(dx)\quad\text{for $i\in\lbrace 0,1\rbrace$ and $n\in\mathbb{N}$}
\end{align}
with a projection operator $P_0$:
\begin{align}
P_0 \vert f_1\rangle=\frac{\langle f_1\vert f_0\rangle}{\langle f_0\vert f_0\rangle}\vert f_0\rangle=\sqrt{\tilde{q}}\vert f_0\rangle
\end{align}
this leads to the conditional expectation moments
\begin{align}
\mathbb{E}[H_{k_i}^{(q)}(x_1)\vert x_0]=\tilde{q}^{k_i/2}H_{k_i}^{(q)}(x_0)
\label{ceH}
\end{align}
since $\mathbb{E}[\vert f_1\rangle^{\otimes n}\vert \vert f_0\rangle]=(P_0 \vert f_1\rangle)^{\otimes n}$ and $H_{n}^{(q)}(x_i)\vert 0\rangle=\vert f_i\rangle^{\otimes n}$ ($H_{n}^{(q)}(x_i)$ is the normal ordering of $x_i^n$ see section \ref{sec:2}).

Conditional expectation serves as a transition operator projecting states from $f_1$ to $f_0$. This operator has an integral representation:
\begin{align}
\mathbb{E}[x_1^{k_i}\vert x_0]=\int y^{k_i}k(x_0,dy) 
\end{align}
with $x_1=x_0+y$ and $k(x,dy)$ being the transition probability,  defined as:
\begin{align}
k(x,dy)=p_{\sqrt{\tilde{q}}}(x,y)\nu_q(dy)
\label{tran}
\end{align}
where $\nu_q$ is defined in \eqref{densitySYK}. Moreover, from \eqref{ceH} we know that $p_r(x,y)$ is the kernel of the $q$-Hermite polynomial:
\begin{align}
p_r(x,y):=\sum_{n=0}^{\infty}\frac{r^n}{[n]_q!}H_n^{(q)}(x)H_n^{(q)}(y)
\end{align} 
This transition probability is sometimes referred to as the conditional $q$-normal distribution, with its moments provided in \cite{Szab}. 
Expanding the monomial $x_0^{k_i}$ into $q$-Hermite polynomials \eqref{top}, its conditional expectation can be expressed as
\begin{align}
\mathbb{E}[x_1^{k_i}\vert x_0]=\sum_{m_i=0}^{\lfloor k_i/2\rfloor}c_{m_i,k_i}\sqrt{\tilde{q}}^{k_i-2m_i}H_{k_i-2m_i}^{(q)}(x_0).
\label{ee}
\end{align} 
This expression recovers \eqref{stoch}. For further details on the $q$-OU process
and its associated transition probability, please refer to Theorem 4.6 of \cite{Bozejko}.

To summarize, the computation of a Majorana fermion product within the $i$-th interval, also referred to as a moment calculation building block of equation \eqref{moments1} (an elementary element of the substructure $\mathcal{B}$), is effectively realized by the conditional moments of a $q$-OU process in the chord Hilbert space. This can be expressed as 
\begin{align*}
\Psi_{\sum_{m=1}^{i-1}\alpha_{k_m}+1}\dots\Psi_{\sum_{m=1}^i\alpha_{k_m}}\longrightarrow \mathbb{E}[x_1^{k_i}\vert x_0]
\end{align*}
which depends on the variable $x_0$.  As a result,  $B(z)$  becomes a
bivariate generating function defined as
\begin{align*}
B(z,x_0)=\theta z\sum_{k=0}^{\infty}
\mathbb{E}[x_1^{k}\vert x_0]z^k
\end{align*}
Here $\theta z$ is used to mark each building block (we have redistributed the $n-2j$ $\theta$ in \eqref{moments1} so that each interval has one $\theta$). See section $3$ of \cite{Wu} for the detailed explanation.
The explicit formula is given by:
\small
\begin{align}
B(z,x_0)=\theta\sqrt{1-q}\sum_{k\geq 0}\frac{\sqrt{\tilde{q}(1-q)}^k H_{k}^{(q)}(x_0)}{[k]_q!}\sum_{j\geq0}(-1)^j\frac{[k+j]_q!}{[j]_q!}q^{j(j+1)/2}\left( \frac{1-\sqrt{1-4z^2/(1-q)}}{2z/\sqrt{1-q}}\right)^{2j+k+1}
\end{align}
\normalsize
it takes the form of a continued fraction:
$$
B(z,x_0)=\frac{\theta z}{1-\sqrt{\tilde{q}}x_0z-\frac{(1-\tilde{q})[1]_q z^2}{1-\sqrt{\tilde{q}}q x_0z-\frac{(1-\tilde{q}q)[2]_q z^2}{1-\sqrt{\tilde{q}}q^2x_0z-\frac{(1-\tilde{q}q^2)[3]_q z^2}{1-\sqrt{\tilde{q}}q^3x_0 z-\frac{(1-\sqrt{q}q^3)[4]_q z^2}{1-\cdots}}}}}.
$$
This is due to the transition probability $k(x_0,dy)$ representing the measure of an orthogonal polynomial \cite{Szab}:
\begin{align*}
P_{n+1}(y\vert x_0,\sqrt{\tilde{q}},q)=(y-\sqrt{\tilde{q}}q^nx_0)P_{n}(y\vert x_0,\sqrt{\tilde{q}},q)-(1-\tilde{q}q^{n-1})[n]_qP_{n-1}(y\vert x_0,\sqrt{\tilde{q}},q)
\end{align*}
The remaining random variable $x_0$ in $B(z,x_0)$ corresponds to the singletons, which need to be paired out. Consequently, 
the moment formula can be obtained from 
\begin{align}
m_n&=[z^n]\langle 0\vert z\frac{d}{dz} \log\frac{1}{1-B(z,x_0)}\vert 0\rangle\nonumber\\
&=n[z^n]\langle 0\vert  \log\frac{1}{1-B(z,x_0)}\vert 0\rangle
\label{gm}
\end{align}
Here again, the vacuum expectation corresponds to the expectation value with respect to the $q$-Gaussian distribution
$$
\langle 0 \vert .\vert 0\rangle=\int . \nu_q(x_0)dx_0
$$
with $\nu_q$ defined in \eqref{densitySYK}.
Therefore, combining the above information, we can derive \eqref{mm} from \eqref{gm}. See appendix \ref{appd:3} for the detailed calculation.

It is clear that our moment calculation is governed by the parameter \(\tilde{q}\), where \(0\leq\tilde{q}\leq1\). We therefore conclude this section by examining two extreme cases:
\begin{itemize}
\item
when $\tilde{q}=0$, 
$$
\mathbb{E}[x_1^{k}\vert x_0]=\int x_0^{k}\nu_{q}(x_0)dx_0= \left\{\begin{array}{ll}c_{k/2,k} \quad\quad & \text{for $k$ even}  \\ 0 \quad & \text{for $k$ odd} \end{array}\right.
$$
with $c_{m,k}$ is defined in \eqref{opc}. The corresponding generating function is reduced to 
\begin{align*}
B(z,x_0)=B(z)&=\theta\sqrt{1-q}\sum_{j\geq0}(-1)^j q^{j(j+1)/2}\left( \frac{1-\sqrt{1-4z^2/(1-q)}}{2z/\sqrt{1-q}}\right)^{2j+1}\\
&=\theta  \sum_{k\geq0} z^k\int x_0^{k}\nu_{q}(x_0)dx_0 
\end{align*}
the moment becomes:
$$
m_n=n[z]^n\log\frac{1}{1-B(z)}
$$
This recovers the results of the case $c=1$, see equation $(9)$ of \cite{Wu}.
\item when $\tilde{q}=1$, the conditional expectation become trivial 
$$
\mathbb{E}[x_1^{k}\vert x_0]=\sum_{m_0}^{\lfloor k/2\rfloor}c_{m,k}H_{k-2m}^{(q)}(x_0)=x_0^k
$$ 
$B(z,x_0)$ becomes
$$
B(z,x_0)=\frac{\theta z}{1-x_0 z}
$$
The $m_n$ is nothing more than a shift of the moment of the $q$-Gaussian distribution by $\theta$:
\begin{align*}
m_n&=n[z^n]\langle 0\vert  \log\frac{1}{1-B(z,x_0)}\vert 0\rangle\\
&=n[z^n]\langle 0\vert \sum_{k\geq 0}
\frac{\sum_{i=0}^{k-1}\binom{k}{i}x_0^i \theta^{k-i}}{k}z^k
\vert 0\rangle\\
&=\langle 0  \vert \sum_{i=0}^{n-1} \binom{n}{i}x_0^i \theta^{n-i}\vert 0\rangle\\
&=\int (x_0+\theta)^n \nu_q dx_0-\mathbb{E}[x_0^n]
\end{align*}
\end{itemize}
\section{Mixtures of different independences}
\label{sec:5}
In non-commutative probability theory, various forms of independence have been studied, analogous to the concept of independence in classical probability theory, but applied to non-commutative random variables. Naively speaking, these notions of independences tells us how to calculate mixed moments involving different random variables. One of the earliest and most famous notions is the concept of freeness, which was originally introduced by Voiculescu \cite{Voiculescu} in the context of algebra calculation. Later, Speicher \cite{Speicher6} contributed by exploring the combinatorial facets of freeness, particularly in the realm of non-crossing partitions, through the development of the moment-cumulant formula for free random variables. In our work, we connect our moment computation to a definition of a mixture of two types of independence from the combinatorial aspects, inspired by Speicher's insights.

In the context of our moment calculation in the perturbative SYK model, what we are actually calculating is the sum over all possible joint moments of $D_c$ and $H_{\text{SYK}}$. These joint moments can be effectively expressed using joint cumulants, which depend on specific set partitions. This implies a coexistence of the two forms of independence characterizing our model. The notion of independence discussed here refers to the interdependence between different variables, which essentially tells us how to calculate their mixed moments in terms of their individual moments. In the following section, we introduce a moment-cumulant relation to define this amalgamation of independences between $D_c$ and $H_{\text{SYK}}$. To develop a calculation approach for our mixed moments, we draw inspiration from Speicher's concept of a ``universal product" \cite{Speicher3} and the notion of $\varepsilon$-independence introduced by M\l otkowski \cite{Mlotkwski}.

Let's first define a lineal functional $\varphi$ as
\begin{align}
\varphi:=\text{tr}\otimes E
\end{align}
with $E$ is the expectation over the random coupling $J$, thus
$$
\varphi (H_{\text{SYK}}^k)=<\text{tr}H_{\text{SYK}}^k>_J\quad\text{and}\quad \varphi (D_c^l)=\text{tr} D_c^l
$$
Since $D_c$ is a constant diagonal matrix, it can be approximated by a random variable, i.e. its eigenvalue distribution converges to that of the same random variable. For further elaboration, see section 3.5 of \cite{Speicher2}. On the other hand, $H_{\text{SYK}}$ is characterised by a distribution that converges to a $q$-Gaussian variable. This implies different forms of independence between $D_c$ and $H_{\text{SYK}}$. These two different types of variables are distinguished by their respective moment-cumulant relationships. In the context of combinatorial structure, the free variable involves counting only non-crossing partitions, while the other involves the inclusion of crossing terms within partitions. We denote the convergences by \footnote{We adapt the notion of convergence from Speicher's lecture notes \cite{Speicher2}: we say that $a$ converges to $b$ if every moment in $a$ converges to the corresponding moment in
$b$.}
\begin{align*}
&H_{\text{SYK}}\xrightarrow{\text{dist}}x \quad\Leftrightarrow\quad \varphi(H^k_{\text{SYK}})\xrightarrow{\text{converge}}\varphi(x^k)\\
&\theta D_c\xrightarrow{\text{dist}}d \quad\Leftrightarrow \quad\frac{1}{r}\varphi((\theta D_c)^k)\xrightarrow{\text{converge}}\varphi(d^k)
\end{align*}
with $x$ is a $q$-Gaussian element and $d$ is a free variable with moments $\varphi (d^n)=\theta ^n$.
Thus, we have two different moment-cumulant relations for $\varphi(x^n)$ and for $\varphi(d^n)$.
\begin{align}
\varphi(x^{2n})=\sum_{\pi\in \mathcal{P}_2(2n)} q^{\text{cr}(\pi)}\prod_{i=1}^n k_2(x)
\label{m-c1}
\end{align}
the sum above is over all the perfect matching of $[2n]$, therefore $\varphi(x^k)=0$ for $k$ odd. $k_2(x)$ is the second combinatorial cumulant associated with the $q$-Gaussian distribution and $k_j(x)=0$ for $j>0$. For the detailed definition of the combinatorial cumulant, see section $6.2$ of \cite{ANSHELEVICH1}. Worth noting is that for a normalized centred $q$-Gaussian, $k_2(x)=1$, this particular moment-cumulant relationship subsequently  recovers the famous Riordan-Touchard formula. Here are few examples: $\varphi(x^2)=k_2(x)$, $\varphi(x^4)=(2+q)k^2_2(x)$, $\varphi(x^6)=(5+6q +3q^2+q^3)k^3_2(x)$.

For $d$, since it is a free random variable, the moment-cumulant relation of free probability theory is aptly applied:
\begin{align}
\varphi(d^{n})=\sum_{\pi\in NC(n)}\prod_ {B_i\in\pi}k_{\vert B_i\vert}(d)
\label{m-c2}
\end{align}
where the sum is over all the non-crossing partition of $[n]$. $B_i$ signifies the constituent blocks of a partition denoted as $\pi=\lbrace B_1,B_2,\cdots,B_s\rbrace$ with each block $B_i$ defined as $B_i=(l_1,l_2,\cdots l_{\vert B_i \vert})$, wherein $l_j \in [n]$ for $1 \leq j \leq \vert B_i \vert$. The parameter $\vert B_i \vert$ denotes the number of elements within the block $B_i$. Moreover, $k_{j}(d)$ denotes the $j$-th free cumulant  of $d$. Here are few examples: $\varphi(d)=k_1(d)$, $\varphi(d^2)=k_2(d)+k^2_1(d)$, $\varphi(d^3)=k_3(d)+3k_1(d)k_2(d)+k^3_1(d)$ and $\varphi(d^4)=k_4(d)+4k_1(d)k_3(d)+6k^2_1(d)k_2(d)+k^4_1(d)$.

It should be noted that, for simplicity's sake, we don't use subscripts to distinguish between the two different types of cumulant. Instead, we use different variables to distinguish them: for a variable denoted as $d$, the corresponding $j$-th cumulant $k_j(d)$ is of the free type; whereas for a variable denoted as $x$, the corresponding $j$-cumulant $k_j(x)$ is a combinatorial cumulant, this type of cumulant depending on the partition and taking into account any associated crossings within that partition.

First, to compute the $n$-th joint moments $\varphi (a_1a_2\cdots a_n)$ with $a_j\in\lbrace x,d \rbrace$, $\forall j\in[n]$, we need to define a partition to divide the sequence $\lbrace a_1, a_2,\cdots,a_n\rbrace$ into different blocks so that every element in each block belongs to the same subset (all the $a$'s in a block are either $x$ or $d$). We denote the index set $\mathbf{i}=\lbrace i_1,i_2,\cdots,i_n\rbrace$ and assign each $a_j$ a decoration index $i_j$ which indicates to which subsets $a_j$ belongs.
$$
i_j=\left\{\begin{array}{ll}1 \quad\quad & \text{for $a_j=x$}  \\ 2 \quad & \text{for $a_j=d$} \end{array}\right.
$$
Since we have only two different values in $\mathbf{i}$, the kernel $\ker \mathbf{i}$ is a partition with two blocks, we call them $V$ with decoration index $1$ and $W$ with decoration index $2$. If the sequence $\lbrace a_1,a_2,\cdots,a_n\rbrace$ has $m$ elements with decoration index $1$, then $V=( v_1,v_2,\cdots, v_m)$ and $W=( w_1,w_2,\cdots,w_{n-m})$ where $v_1,v_2,\cdots,v_m$ and $ w_1,w_2,\cdots,w_{n-m}$ are both elements in $[n]$.
In mixed moment computation, any permissible partition $\pi$ must be a subset of $\ker\mathbf{i}$, indicated by $\pi\leqslant \ker\mathbf{i}$, \.i.e. each block in $\pi$ must be fully contained within a block of $\ker\mathbf{i}$. This implies that two elements $j$ and $k$ share a block in $\pi$ if and only if their indices satisfy $i_j=i_k$.


Based on the above information, 
we propose a joint moment-cumulant relation such that the mixed moments of $H_{\text{SYK}}$ and $Dr$ can be approximated by 
\begin{align}
\varphi (a_1a_2\cdots a_n)&=\sum_{\substack{\pi=\pi_1 \cup\pi_2\\\text{$\pi_1\in\mathcal{P}(V)$, $\pi_2\in\mathcal{NC}(W)$} \\ \ker \mathbf{i}=\lbrace V,W\rbrace}}\tilde{q}^{\text{br}(\pi)} q^{\text{cr}(\pi_1)}\prod_{B_i\in\pi_1}k_{\vert B_i\vert} \prod_{B_j\in \pi_2} k_{\vert B_j\vert}\nonumber \\
&=
\sum_{\substack{\pi=\pi_1 \cup\pi_2\\\text{$\pi_1\in\mathcal{P}(V)$, $\pi_2\in\mathcal{NC}(W)$} \\ \ker \mathbf{i}=\lbrace V,W\rbrace}}\tilde{q}^{\text{bc}(\pi)} q^{\text{cr}(\pi_1)}k_{\pi}(a_1,a_2,\cdots ,a_n)
\label{m-c}
\end{align}
where $\ker\mathbf{i}$ is the corresponding decoration index set of $\lbrace a_1,a_2,\cdots,a_n\rbrace$, $V$ is the subset of $\ker\mathbf{i}$ with decoration index $1$, and $W$ is the subset of $\ker\mathbf{i}$ with decoration index $2$.  The allowed partition $\pi$ in this moment-cumulant relation is shaped by the index set $\mathbf{i}$. Thus, the sums in \eqref{m-c} are over all the unions of two types of partitions: $\pi_1$ for the partition of $V $ and $\pi_2$ for the non-crossing partition of $W$. There are two kinds of crossings that we need to consider in $\pi$. One is inside $V$, we denote the corresponding crossing number $\text{cr}(\pi)$ defined as
$$
\text{cr}(\pi_1):=\vert \lbrace \lbrace k,p,q,l\rbrace\in\pi_1: k<p<q<l \text{ s.t. $k\sim_{\pi_1} q\nsim_{\pi_1} p\sim_{\pi_1} l$}\rbrace\vert
$$
$k\sim_{\pi_1} q$ means $k$ and $q$ are in the same block of $\pi_1$ and $q\nsim_{\pi_1} p$ means $q$ and $p$ are in different blocks of $\pi_1$.
The other type of crossings happens between $\pi_1$ the partition of $V$ and the other subset $W$. The corresponding crossing number is denoted as $\text{bc}(\pi)$ and defined as
$$
\text{bc}(\pi):=\vert\lbrace \lbrace k,q\rbrace\in\pi_1: k<p<q<l \text{ s.t. $k\sim_{\pi_1}q$ and $\lbrace p,l\rbrace\in W$}\rbrace\vert
$$
Note that $p$ and $l$ are in the subset $W$, but they don't have to be in the same block.

The expression $k_{\pi}$ denotes the product of free cumulants aligned with the block structure of partition $\pi$:
\begin{align*}
k_{\pi}(a_1,a_2,\cdots ,a_n)=\prod_{B_i\in\pi}k_{\vert B_i\vert}(a_{i_1},a_{i_2},\cdots ,a_{i_{\vert B_i\vert}}).
\end{align*}
It combines two different types of cumulants, determined by the internal block arrangement in $\pi$.  In particular, a block contained in subset $V$,  contributes a combinatorial cumulant (as defined in \eqref{m-c1}), while a block within $W$ contributes a free cumulant (as defined in \eqref{m-c2}). This resembles a similar scenario found in \cite{Speicher4}. One can also check that all mixed cumulants vanish, \i.e. for $n\geq 2$, $k_{n}(a_{i_1},a_{i_2},\cdots,a_{i_n})=0$ if at least two of the indices in $(i_1,i_2,\cdots,i_n)$ are different. This can be proved using Lemma $6$ in \cite{Ebrahimi}.

Since the sequence $\lbrace a_1,a_2,\cdots,a_n\rbrace\in\lbrace x,d\rbrace^n$, where $x$ is a $q$-Gaussian with only one nonzero cumulant $k_2(x)=1$, it is possible to restrict the partition of the set $V$ to a perfect matching $\mathcal{P}_2(V)$. This restriction gives an expression for the mixed moment as follow:
\begin{align}
\varphi(a_1a_2\cdots a_n)=\sum_{\substack{\pi=\pi_1 \cup W\\\text{$\pi_1\in\mathcal{P}_2(V)$} \\ \ker \mathbf{i}=\lbrace V,W\rbrace}}
\tilde{q}^{\text{bc}(\pi)}q^{\text{cr}(\pi_1)}\varphi(\prod_{j\in W}a_j)
\label{j-m}
\end{align}
Note that in \eqref{j-m} the rearrangement of the elements within the set $V$ is done by $\pi_1$, while the elements in $W$ remain unchanged. This selective treatment stems from the recognition that the definitions of the two different types of crossing only affect the elements inside the set \(V\) when we first partition the elements inside \(V\). After $\pi_1$, the redistribution, all the contributions of the crossings have already been counted, we can simply put all the elements of $W$ together and calculate the expectation of the product of the corresponding $a_j$.
Furthermore, the challenge of expressing the mixed moments as products of individual moments of variables $x$ and $d$ is highlighted by equation $\eqref{j-m}$, mainly due to the introduction of the parameter $\tilde{q}$. This parameter enumerates all blocks within the subset $V$ that intersect with the complementary subset $W$. Since it concerns only the blocks in $V$,  it can be understood as a $\tilde{q}$-deformation of the original moments of $x$. This interpretation is consistent with the $q$-OU process described in section \ref{sec:4}, where an original $q$-Gaussian variable transforms into another $q$-Gaussian variable via a stationary Markov transition. Consequently, moments under this transition become conditional expectation moments for the new variable. This established joint moment-cumulant relation underlines the creation of a new independent relations by the $q$-OU process.

In conclusion, as evidenced by the equations $\eqref{m-c}$ and $\eqref{j-m}$, we observe a one-parameter controlled mixture of independent relations. This arrangement endows the elements within the subset $V$ with a distinct $\tilde{q}$-independent character that interpolates between classical independence and Boolean independence.
For a given $q$, if $\tilde{q}=0$,  equation \eqref{j-m} is nonzero only when any block within $\pi_1$, the partition of the subset $V$, does not intersect with the subset $W$. Therefore we can restrict $\pi_1$ to an interval partition $\mathcal{I}(V)$. In this case, elements in the subset $V$ become Boolean independent. The mixed moment formula becomes:
\begin{align*}
\varphi (a_1a_2\cdots a_n)
=\sum_{\substack{\pi_1\in\mathcal{I}(V) \\ \ker \mathbf{i}=\lbrace  V,W\rbrace}}\prod_{B_i\in\pi_1}\varphi( \prod_{j \in B_i}a_j)\varphi(\prod_{k\in W}a_k)
\end{align*}

When $\tilde{q}=1$, with given $q$, $a_j$ commute with $a_k$ when $a_j\in V$ and $a_k\in W$. Therefore, equation \eqref{j-m} factorize the partitioned expectations into two blocks $V$ and $W$, in this case, elements in $V$ and elements in $W$ are classical independent to each other (they commute):
$$
\varphi (a_1a_2\cdots a_n)=\varphi( \prod_{j \in V }a_j)\varphi(\prod_{k \in W }a_k)
$$
When $q=\tilde{q}=0$, \eqref{m-c} recovers to the joint moment-cumulant relation in free probability yet still under the restriction of the decoration index $\mathbf{i}$:
$$
\varphi (a_1a_2\cdots a_n)=
\sum_{\substack{\pi \in NC(n)\\ \pi\leq\ker\mathbf{i}}}k_{\pi}(a_1,a_2,\cdots a_n)
$$
with $NC(n)$ is the set of all non-crossing partitions of $[n]$. $\pi\leq\ker\mathbf{i}$ means each block of $\pi$ contains in one of the blocks of $\ker\mathbf{i}$. The partition cumulant $k_{\pi}$ here factorizes in a product according to the block structure of
$\pi$. When $q=\tilde{q}=1$, equation \eqref{m-c} becomes the classical moment-cumulant relation with decoration index $\mathbf{i}$ restriction:
$$
\varphi (a_1a_2\cdots a_n)=
\sum_{\substack{\pi \in \mathcal{P}(n)\\ \pi\leq\ker\mathbf{i}}}k_{\pi}(a_1,a_2,\cdots a_n)
$$
$\mathcal{P}(n)$ here is the set of all the partition of $[n]$ and $k_{\pi}$ now becomes the product of classical cumulants according to the block structure of $\pi$.

We end this section with the case $n=4$ as an example to demonstrate how to calculate the joint moment from equation \eqref{m-c} and compare it with the results from equation \eqref{mm}.  In this scenario, the sequence $(a_1, a_2, a_3, a_4)$ has a total of 8 possibilities.  Each element $a_j$ can be either $x$ or $d$, with $j$ in the range $[4]$. These arrangements consist of $(x, x, x, x)$, $(x, x, d, d)$, $(x, d, x, d)$, $(x, d, d, x)$, $(d, x, x, d)$, $(d, x, d, x)$, $(d, d, x, x)$, and $(d, d, d, d)$. We emphasise again that $x$ is a $q$-Gaussian with only one  nonzero cumulant  $k_2(x)=1$. Thus each block in $\pi_1$ contains only $2$ elements.  We compute the corresponding joint moment of the above sequences by using \eqref{m-c}:
\begin{itemize}
\item
$(x,x,x,x)$: $\mathbf{i}=\lbrace 1,1,1,1\rbrace $ then $\ker\mathbf{i}=\lbrace (1,2,3,4)\rbrace$. The allowed partitions $\pi=\pi_1\in\mathcal{P}_2(4)$ can be either: $\lbrace (1,2),(3,4)\rbrace$, $\lbrace (1,4),(2,4)\rbrace$ and $\lbrace (1,3),(2,4)\rbrace$. for the last partition, we have a crossing inside $V$, therefore $$\varphi(x^4)=2k_2^2(x)+qk_2^2(x)$$.
\item
$(x,x,d,d)$: $\mathbf{i}=\lbrace 1,1,2,2\rbrace $ then $\ker\mathbf{i}=\lbrace (1,2),(3,4)\rbrace$. The allowed partitions $\pi=\pi_1\cup\pi_2$ with $\pi_1=\lbrace (1,2)\rbrace$ and $\pi_2$ can be either: $\lbrace (3,4)\rbrace$ or $\lbrace (3),(4)\rbrace$. There is no crossing in these partitions, we get $$\varphi(x^2d^2)=k_2(x)k_2(d)+k_2(x)k_1^2(d)=\varphi(x^2)\varphi(d^2).$$
Same results for $\varphi(xd^2x)$, $\varphi(dx^2d)$ and $\varphi(d^2x^2)$ because of the cyclic invariant of $\varphi$.
\item  
$(x,d,x,d)$: $\mathbf{i}=\lbrace 1,2,1,2\rbrace $ then $\ker\mathbf{i}=\lbrace (1,3),(2,4)\rbrace$. The allowed partitions $\pi=\pi_1\cup\pi_2$ with $\pi_1=\lbrace (1,3)\rbrace$ and $\pi_2$ can be either: $\lbrace (2,4)\rbrace$ or $\lbrace (2),(4)\rbrace$. There is a crossing between $V$ and $W$ in both partitions, therefore
 $$\varphi(xdxd)=\tilde{q}(k_2(x)k_2(d)+k_2(x)k_1^2(d))=\tilde{q}\varphi(x^2)\varphi(d^2).$$
Same results for $\varphi(dxdx)$. One can check that this coincides with the calculation in Appendix \ref{appd:2}.
\item
$(d,d,d,d)$: $\mathbf{i}=\lbrace 2,2,2,2\rbrace $ then $\ker\mathbf{i}=\lbrace (1,2,3,4)\rbrace$. The allowed partitions $\pi=\pi_2\in NC(4)$: $\lbrace (1,2,3,4)\rbrace$, $\lbrace (1,2,3),(4)\rbrace$,$\lbrace (1,2,4),(3)\rbrace$, $\lbrace (1,3,4),(2)\rbrace$, $\lbrace (1),(2,3,4)\rbrace$, $\lbrace (1,2),(3,4)\rbrace$, $\lbrace (1,4),(2,4)\rbrace$, $\lbrace (1,2),(3),(4)\rbrace$, $\lbrace (1,3),(2),(4)\rbrace$, $\lbrace (1,4),(2),(3)\rbrace$, $\lbrace (1),(2,3),(4)\rbrace$,  $\lbrace (1),(2,4),(3)\rbrace$, $\lbrace (1),(2),(3,4)\rbrace$ and $\lbrace (1),(2),(3),(4)\rbrace$. Therefore
$$ \varphi(d^4)=k_4(d)+4k_1(d)k_3(d)+2k_2^2(d)
+6k_1^2(d)k_2(d)+k_1^4(d)$$
Note that in this case, $\lbrace(1,3),(2,4)\rbrace$ is the only partition in $\mathcal{P}(4)$ is not allowed, this is because no crossing is allowed inside $W$.
\end{itemize}
By summing all the $8$ possibilities above, we get the $4$-th moment of $x+d$
\begin{align*}
\varphi\left((x+d)^4\right)&=\varphi(x^4)+(4+2\tilde{q})k_2(x)(k_2(d)+k_1^2(d))+\varphi(d^4)\\
&=\varphi(x^4)+(4+2\tilde{q})\varphi(x^2)\varphi(d^2)+\varphi(d^4)
\end{align*}
one can check that with $\varphi(x^4)=2+q$, $\varphi(x^2)=1$ and $\varphi(d^n)=\theta^{n}$, this recover the results in \eqref{mm} with $n=4$.
\section{Numerical results}
\label{sec:6}
We compare our analytical expression \eqref{mm} to numerical calculations of the ensemble averaging moments of \eqref{hamiltonian} at finite $N$ and fixed $p=4$.  Although all expressions derived in this paper are subject to the double-scaling limit \eqref{double-scaled}, they still provide accurate approximations to the exact results for fixed $p\ll N$. The rationale behind this lies in the remarkable consistency between  double-scaling limit analytical expressions and the exact numerical results for finite, fixed $p$ in the SYK model, even for low $N$ values \cite{Garcia1}. In addition, we present a numerical proof of a phase transition present in the spectral density by observing changes in the spectrum with changes in the values of $\theta$ and $c$. When $c=1$, this phase transition is studied in \cite{Wu}. A precise analytical prediction is provided by the analytical combinatorial method. This phase transition also occurs in additive deformed random matrices \cite{Peche}, a free probabilistic interpretation is given in the Appendix \ref{appd:1}. It should be noted that although free probability theory cannot be applied in the DSSYK model, because $H_{\text{SYK}}$ and $D_c$ are not free to each other, they have an additional non-commutative relation depending on $\tilde{q}$, as outlined in the previous sections. Similar spectrum behaviour suggests that a more generalized convolution theory may be required to provide an analytical solution to this phase transition.

As we are comparing our analytical expressions with numerical results for cases with finite $N$ and finite $p$, we must substitute all occurrences of $q$ in \eqref{mm} with its finite $p$ and $N$ rendition, denoted as $q_n(p,n)$ and defined as
\begin{align}
q_n(p,n)=\sum_{c=0}^{p}(-1)^{c}\binom{p}{c}\binom{N-p}{p-c}
\label{qbis}
\end{align} 

We use Mathamatica to calculate the eigenvalues of \eqref{hamiltonian} with input $\theta=5$, $c=2^{N/2}/2,2^{N/2}/4,2^{N/2}/8,2^{N/2}/16$  and $50$ samplings. The corresponding $\tilde{q}$ are $\tilde{q}(1/2)=q_0$
, $\tilde{q}(1/4)=1/2(q_0+q_1)$, $\tilde{q}(1/8)=1/4(q_0+2q_1)$ and $\tilde{q}(1/16)=1/8(q_0+3(q_1+q+2)+q_3)$ with $q_j$ depending on $c$ and $N$ and its formula is given by \eqref{q_j} (see Appendix \ref{appd:2}).
\begin{figure}[H]
\begin{center}
\includegraphics[scale=0.8]{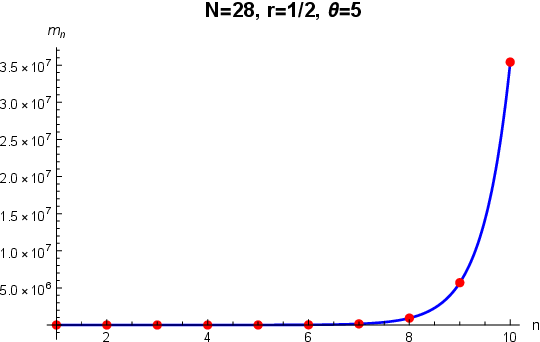}
\includegraphics[scale=0.8]{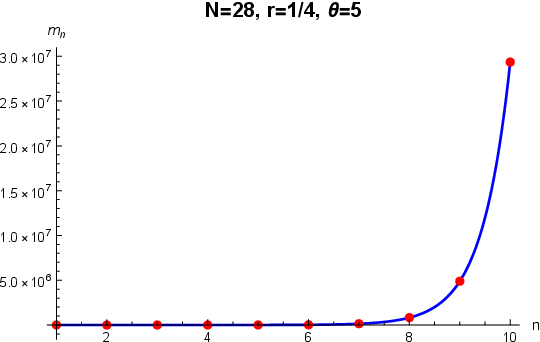}
\includegraphics[scale=0.8]{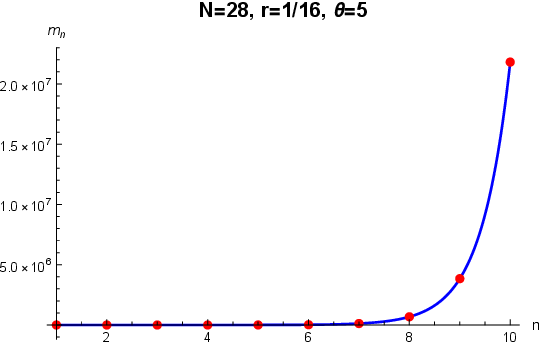}
\includegraphics[scale=0.8]{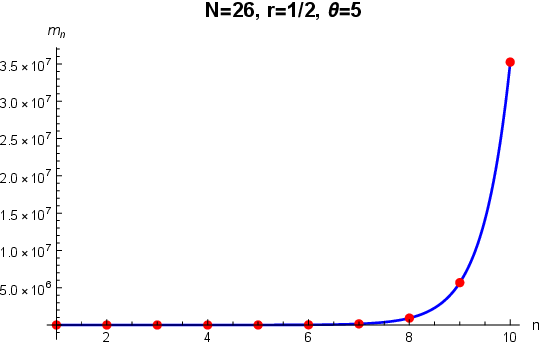}
\includegraphics[scale=0.8]{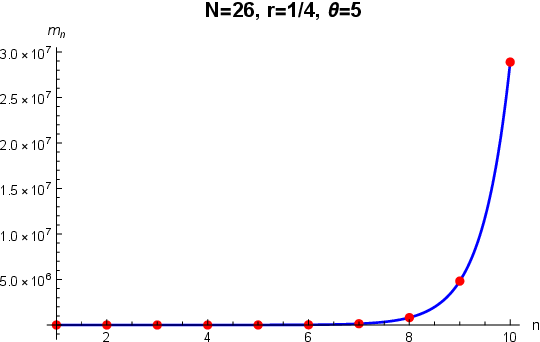}
\includegraphics[scale=0.8]{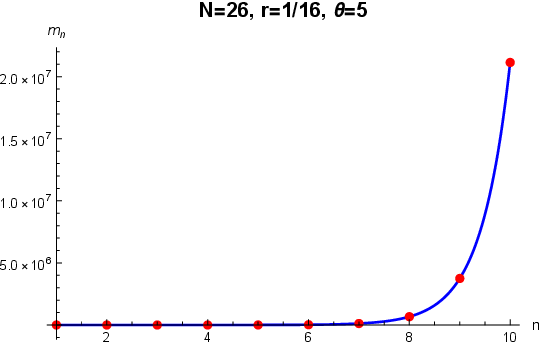}
\caption{Moments calculations comparison: the red dots correspond to the numerical results and the blue lines correspond to our analytical prediction \eqref{mm}.}
\label{num1}
\end{center}
\end{figure}
Figure \ref{num1} displays the first ten ensemble-averaging moments with fixed $\theta$ and various $r$ for $N = 26$ and $N = 28$, respectively. The value of $\theta$ can also be varied. In the following, we provide examples with fixed $r$ for $N=26$.
\begin{figure}[H]
\begin{center}
\includegraphics[scale=0.85]{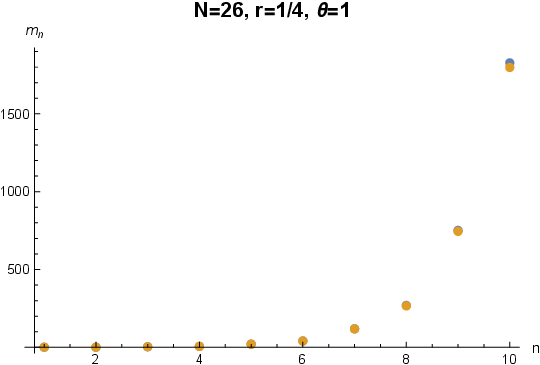}
\includegraphics[scale=0.85]{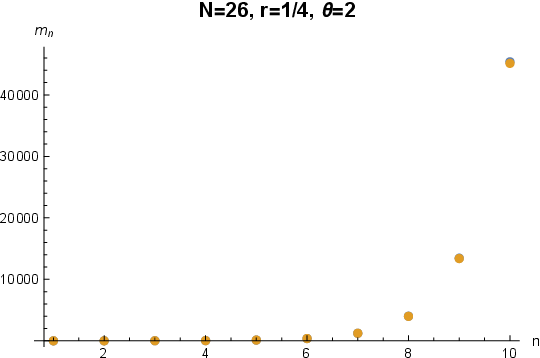}
\includegraphics[scale=0.85]{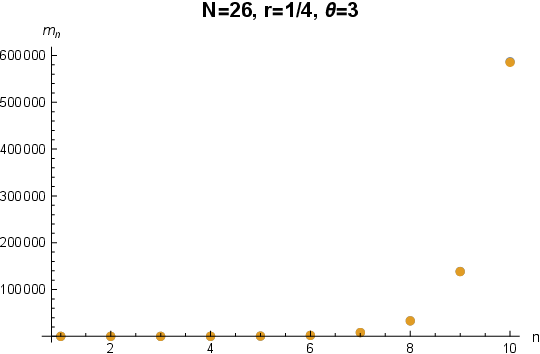}
\includegraphics[scale=0.85]{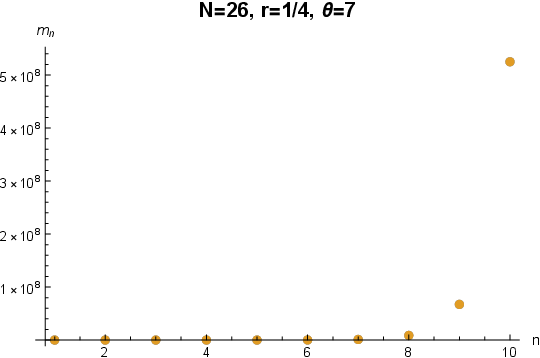}
\caption{Moments calculations comparison with fixed $N$ and $r$: the blue dots correspond to the numerical results and the yellow dots correspond to our analytical prediction \eqref{mm}.}
\label{num2}
\end{center}
\end{figure}
In figure \ref{num2}, the numerical results represented by the blue dots are almost invisible, except for the higher moments. This is because for the first few moments we have excellent agreement between \eqref{mm} and the numerical calculations. As we increase the value of $\theta$, we observe better agreement between the numerical and analytical results. This is somewhat predictable, as the contribution of the constant term in the moment calculation becomes dominant with increasing values of $\theta$.

In general, it can be stated that the moment formula \eqref{mm} is a reliable predictor for any values of $r$ and $c$. However, relying solely on the moment formula does not allow us to comprehend the density of states. For instance, while this formula is applicable for all $r$ and $\theta$, it cannot provide information on the density phase transition that occurs when the spectrum changes from a single to two intervals, with the gap between them depending on the values of $r$ and $\theta$. In the following, we present numerical evidence of the phase transition occurring in the spectrum.
\begin{figure}[H]
\begin{center}
\includegraphics[scale=0.7]{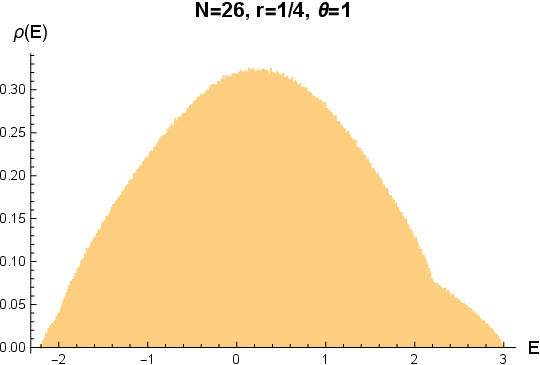}
\includegraphics[scale=0.7]{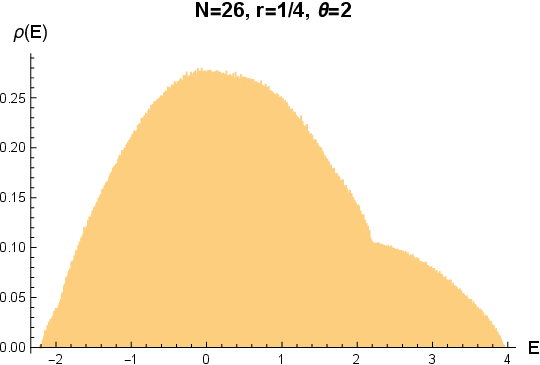}
\includegraphics[scale=0.7]{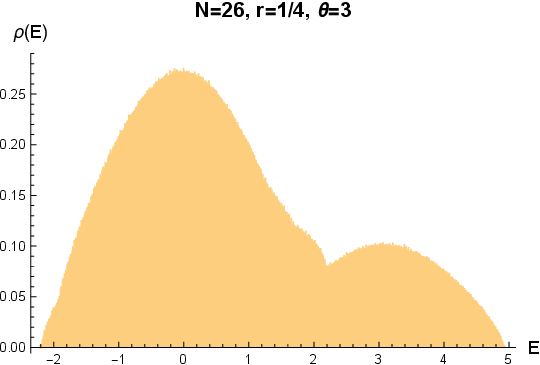}
\includegraphics[scale=0.7]{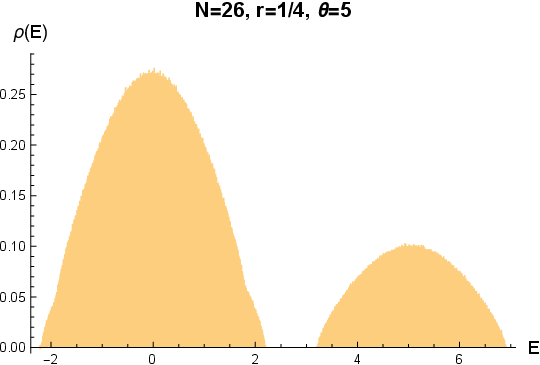}
\caption{Comparison of the numerical spectral density of \eqref{hamiltonian} with $p=4$, $N=26$ and $r=1/4$ for $\theta=1$, $\theta=2$, $\theta=3$ and $\theta=5$ respectively.}
\label{den1}
\end{center}
\end{figure}
Figure \ref{den1} illustrates that as $\theta$ increases while $r$ is held constant, the support of the eigenvalue spectrum changes from a single interval to two disjoint intervals. It clearly exists a critical value of $\theta$.
In the case where $r=2^{-N/2}$ (the $c=1$ case), section 4 of \cite{Wu} demonstrates how to obtain this critical value explicitly by analyzing the singular behaviour of a supercritical functional composition scheme. A similar phase transition occurs within the spectrum of random matrix models, which can be understood by applying the analytic subordination property, valid for the free additive convolution between the random matrices \cite{Capitaine}. Unfortunately, for $r\gg 2^{-N/2}$, the critical value of $\theta$ cannot be computed because the necessary analytical tool is not available due to the complexity of the independence present in this model, as discussed in the previous sections. The other important factor affecting the spectrum is the value of $r$. This is because $r$ determines the distribution of eigenvalues of $D_c$ along with $\theta$. A simple explanation of its significance is that, as $r$ increases while $\theta$ remains fixed, the ``mass" of $D_c$ becomes increasingly influential. This leads to a greater number of eigenvalues of \eqref{hamiltonian} being pulled out from the bulk of the spectrum.   Refer to the illustration provided below for a better understanding.
\begin{figure}[H]
\begin{center}
\includegraphics[scale=.4]{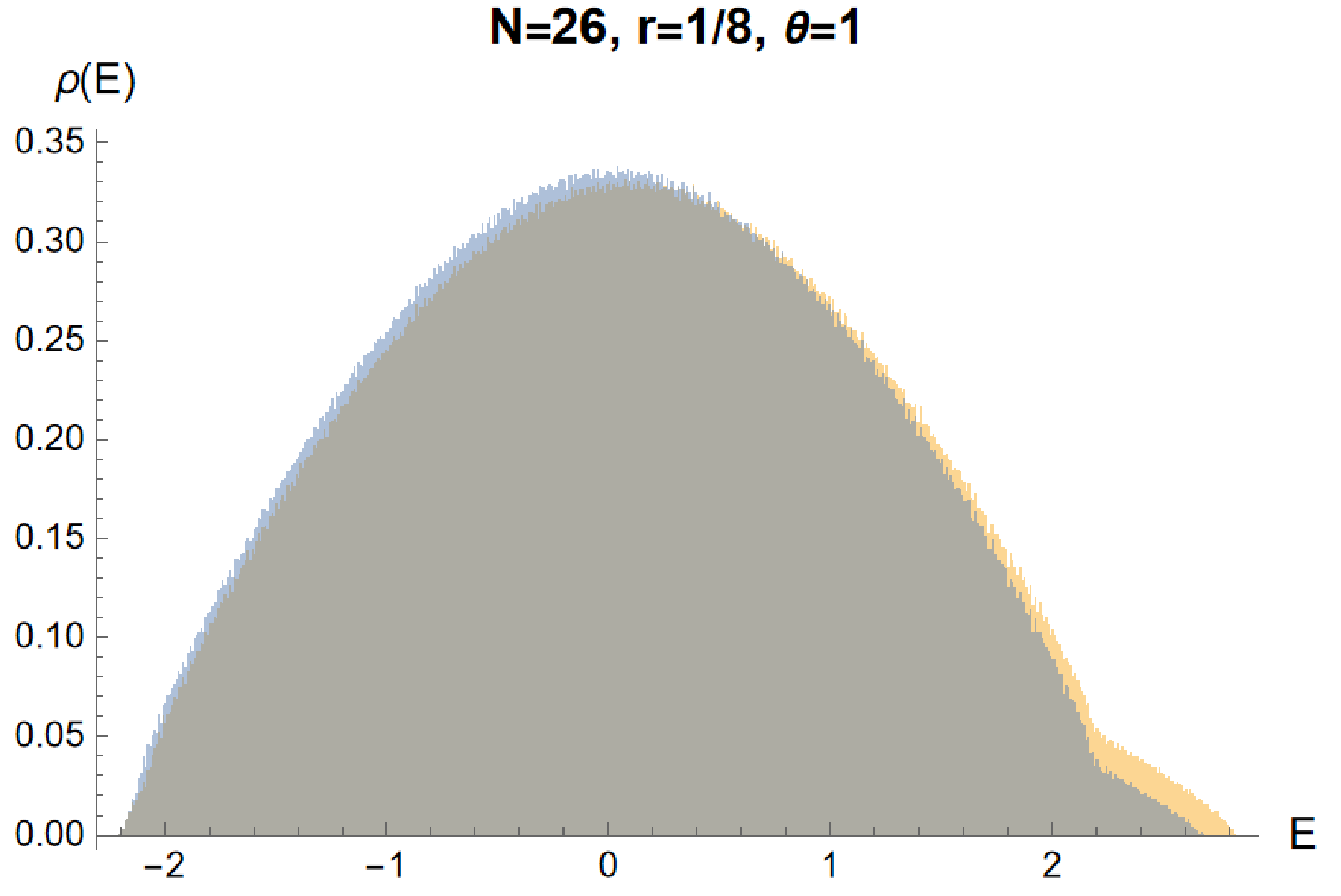}
\includegraphics[scale=.4]{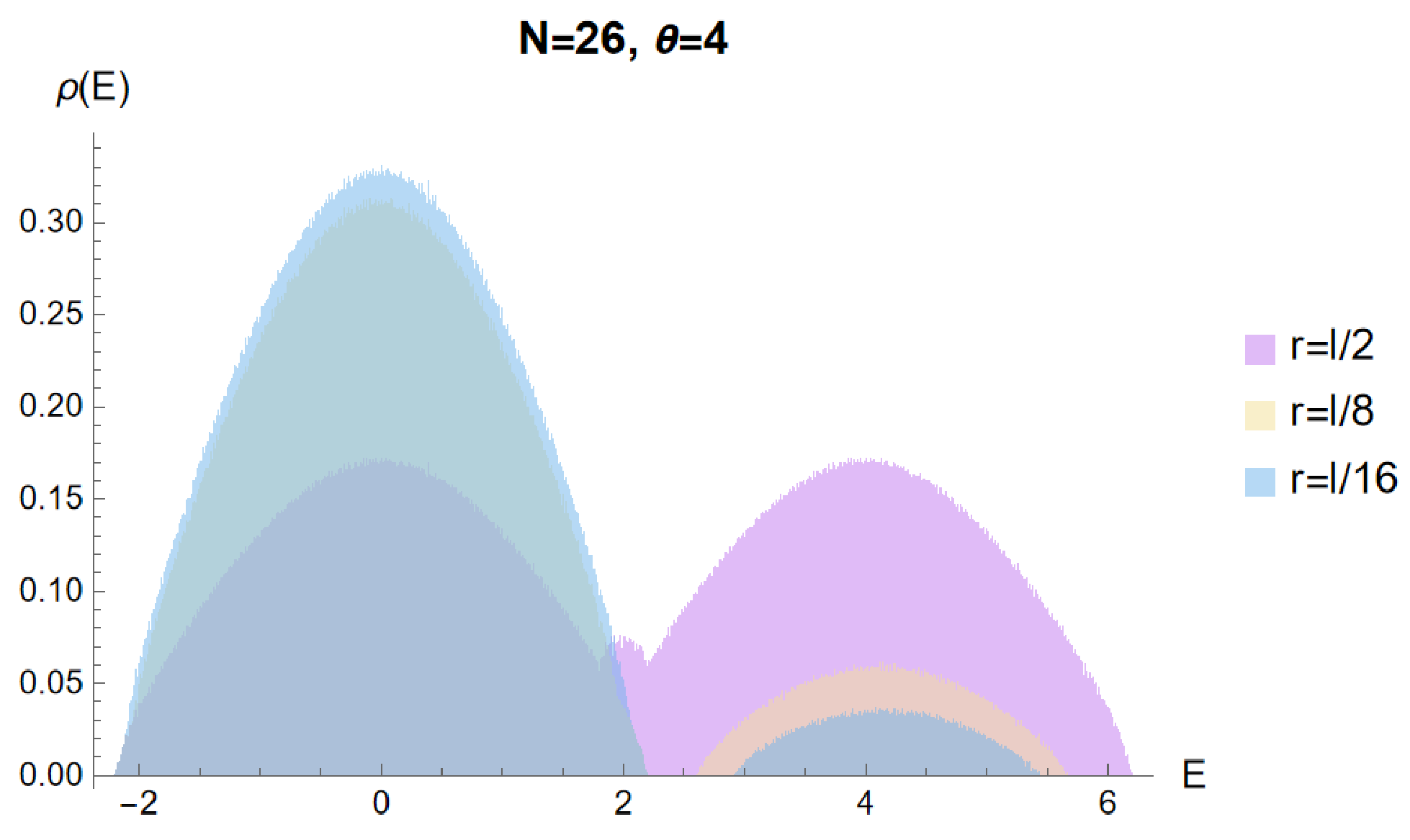}
\caption{Comparison of the numerical spectral density of \eqref{hamiltonian} with $p=4$, $N=26$ for $\theta=1$(top) and $\theta=4$ (bottom) with various values of $r$.}
\label{den2}
\end{center}
\end{figure}
Figure \ref{den2} indicates that increasing $r$ with fixed $\theta$ not only increases the distance between the eigenvalues in \eqref{hamiltonian}, but also determines when the phase transition occurred. $r$ has its own critical value. Consequently, the phase transition in the spectrum of \eqref{hamiltonian} relies on both the value of $r$ and the value of $\theta$.
\section{Conclusion}
\label{sec:7}
We have studied \eqref{hamiltonian}, a SYK model with an extra constant term $D_c$, initially represented as constant diagonal matrices, and later found an equivalent representation using Majorana fermions. Our main achievement is deriving exact expressions for the ensemble-averaged moments of this model. Equation \eqref{mm} for moments is applicable not only to the SYK model with a constant term $D_c$ as defined in equation \eqref{pert} but also to any SYK model perpetuated by a term that has a non-commutative relation with the SYK model, yielding an averaged intersection weight $\tilde{q}$. The particulars of $D_c$ and $\tilde{q}$ do not impact the moments calculation in Section \ref{sec:4}.We observe that, during the moment calculation, the introduction of $D_c$ creates ``walls" that effectively partition the product of SYK Hamiltonians into different intervals.  This partitioning process is characterised by the complex cyclic combinatorial structure \eqref{op_complex}, initially identified in \cite{Wu}. The significance of these ``wall" in the moment calculation lies in their role of splitting the ensemble averaging into two distinct steps. This division is done by introducing the parameter $\tilde{q}$, whose value depends on $r$, the ratio of nonzero entries in $D_c$ and $0\leq\tilde{q}\leq 1$. Consequently, if we consider $\tilde{q}$ (or more precisely $\log\sqrt{\tilde{q}}$) as a time parameter, within each interval separated by ``wall"s, the product of the SYK Hamiltonians effectively transforms into the conditional expectation moments of the $q$-OU process in the chord Hilbert space.  This $\tilde{q}$ controlled interval structure implies the existence of a mixture of different independences within the joint moment calculation. In section \ref{sec:5} we investigated the impact of $D_r$ in the context of non-commutative probability theory. We discovered that the transition in the $q$-OU process induces a $\tilde{q}$ deformation of the initial $q$ Gaussian variables. This deformation alters the commutative and associative independence relations between elements in different intervals, providing an interpolation between Boolean independence at $\tilde{q}=0$ and classical independence at $\tilde{q}=1$. The $\tilde{q}$ joint moment formula proposed in Section \ref{sec:5} differs from the mixed moment involving random and deterministic matrices (see Theorem 6.14 of \cite{Speicher2}). This contrast underscores the special nature of the SYK model, emphasizing that it cannot be regarded as just another random matrix model. To compute the spectral distribution of \eqref{hamiltonian}, a generalized convolution theorem is required to obtain the spectrum of a sum of terms involving a mixture of different independent variables.

\begin{figure}[H]
\begin{center}
\includegraphics[scale=1.]{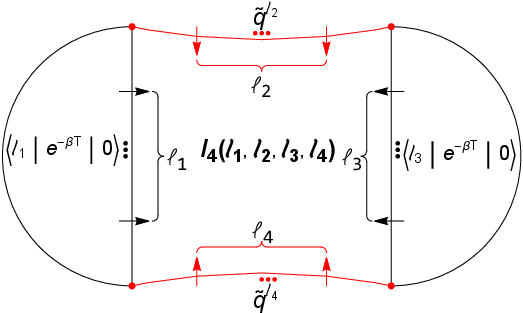}
\caption{Geometry configuration of $Z_4$: The right and left semicircles correspond to two regions of a chord diagram bounded respectively by two disconnected intervals of the chord circle, each contributing to an HH wavefunction, they are connected by two geodesics whose endpoints are inserted operators represented by red points. Each red geodesic contributes a mass action. The arrows represent the outgoing chords, all of which merge in the central region. This region is enclosed by $4$ geodesics of length $l_1$, $l_2$, $l_3$ and $l_4$, it defines an interior state denoted by $I_4(l_1,l_2,l_3,l_4)$.}
\label{wormh}
\end{center}
\end{figure}

Finally, we consider the implications of introducing $D_c$ into the SYK model in relation to gravitational bulk theory. It is important to note that the following suggestions are not rigorous and are intended solely to explore potential directions for future work. If we believe that the DSSYK model has its own picture of bulk gravity \cite{Lin,Berkooz3}, then the introduction of $D_c$ in a sense plays the role of the EOW brane in the Penington-Shenker-Stanford-Yang (PSSY) model \cite{Penington}. Naively speaking, the PSSY model is a simple model that mimics an evaporating black hole, so that the entropic information puzzle is still manifest and the gravitational path integral can be calculated explicitly\footnote{For more recent developments on black holes and the information paradox, see \cite{Almheiri,Penington1,
Almheiri1,Almheiri2}}.  What PSSY  essentially does in the context of free probability\footnote{See \cite{Wang} for more information on interpreting the PSSY model using free probabilities.} is to create a compound Poisson distribution that generalises the Marchenko-Pastur distribution (which is the eigenvalue density of the density matrix of the Page model \cite{Page}) in a particular way by introducing an additional EOW brane term into the Euclidean JT action. Since pure JT gravity is dual to a random matrix model \cite{Saad}, adding EOW branes can be seen as inserting certain operators into the original matrix integral such that the effect of EOW brane actions come out
as gamma function prefactors (see Appendix D of \cite{Penington}). 
The boundary particle formalism used in the PSSY model can be adapted in the chord Hilbert space. Therefore, in the chord Hilbert space, it is possible to explore the potential of constructing a similar model with entanglement entropy following the same pattern as the PSSY model.  To verify this hypothesis, we need to check that all the necessary ingredients are available in the chord Hilbert space. First, consider a partition function in the chord Hilbert space:
\begin{align*}
\langle l_2\vert e^{-\beta T}\vert l_1\rangle=\int_{-\frac{2}{\sqrt{1-q}}}^{\frac{2}{\sqrt{1-q}}}e^{-\beta x}\frac{H_{l_1}^{(q)}(x)H_{l_2}^{(q)}(x)}{\sqrt{[l_1]_q![l_2]_q!}}\nu_{q}(x)dx
\end{align*}
this matrix element measure the probability of propagating through a region with given an incoming state of $l_1$ chords and outgoing state of $l_2$ chords. Additionally, we will use the fact that the SYK model's partition function is a propagator of a particle on $AdS_2$ \cite{Kitaev2}. Hence, $\langle l\vert e^{-\beta T}\vert 0\rangle$ can be interpreted as the Hartle-Hawking wavefunction with $\beta$ denotes the length of the $AdS$ boundary and $l$ represents the length inside the bulk \cite{Okuyama}. 
Next, in PSSY model,  the gravitational path integral has specific boundary conditions, wherein the thermal boundaries on the disk are sandwiched between geodesic trajectories of particle of mass $\mu$. These trajectories contribute a mass action $\mu d$ with $d$ being the geodesic distance. In chord Hilbert space, the geodesic length is discriminated and given by $d=\lambda l$ with $\lambda=-2p^2/N$ and $l$ being the chord number. In order to have a mass action, one can insert a pair of operators into the chord diagram and connect them by a line, the distance between them is determined by the number of chords crossing through this line, each crossing gives a weight factor $\tilde{q}$. Suppose we have $l$ chords crossing through, then the total weight is $\tilde{q}^l$, if $\tilde{q}=q^{\mu}$, we get the mass action term $\tilde{q} ^l=e^{-l\lambda\mu}=e^{-\mu d}$ (see \cite{Lin, Berkooz3} for more details). As explained in section \ref{sec:4}, inserting operators redistributes chords into different intervals, any chord passing through the ``wall" bounded by inserted operators gives a weight contribution $\tilde{q}$.  The length of the interval is determined by how many chords going through. 
Let's assume that there are $n$ intervals on the circle of the chord diagram. Each interval is propagated by an HH wave function $\langle l_i\vert e^{-\beta T}\vert 0\rangle$, and they are separated by a geodesic with length $l_{i+1}$.  Consequently, we have $2n$ outgoing chords extending into the bulk of the chord diagram. Matching all the outgoing chords from different intervals is necessary to connect all these regions. This can be achieved by using the linear coefficients of the product of $q$-Hermite polynomial \eqref{coeffH}
\begin{align}
I_{2n}(l_1,l_2,\cdots l_{2n})=\int_{-2\sqrt{1-q}}^{2\sqrt{1-q}} dE\nu_q(E)\prod_{i=1}^{2n}\frac{H_{l_i}^{(q)}(E)}{\sqrt{[l_i]_q!}}
\label{Iint}
\end{align}
$I_{2n}(l_1,l_2,\cdots l_{2n})$ can be understood as an interior state connecting $2n$ geodesic of fixed regularized lengths $\lbrace l_i\rbrace_{1\leq i\leq2n}$. It is an entangled states of all these $2n$ subregions.
The partition function of the system is then given as
\begin{align}
Z_n=\sum_{l_1\geq0}\cdots\sum_{l_{2n}\geq0}I_{2n}(l_1,l_2,\cdots,l_{2n})\langle l_1\vert e^{-\beta T}\vert 0\rangle\frac{\tilde{q}^{l_2}}{\sqrt{[l_2]_q!}}\cdots \langle l_{2n-1}\vert e^{-\beta T}\vert 0\rangle\frac{\tilde{q}^{l_2}}{\sqrt{[l_{2n}]_q!}}
\label{npf}
\end{align}
This is an analogue of the JT gravity path integral eq $(2.32)$ in \cite{Penington}. See figure \ref{wormh} as an example to illustrate the geometry configuration of equation \eqref{npf}. For simplicity, we won't draw nodes and chords below.
By using the properties of the $q$-Hermite polynomial, on can further simplify $Z_n$
\begin{align}
&Z_n=\int_{-2/\sqrt{1-q}}^{2/\sqrt{1-q}}y(E)^n \nu_q(E)dE\text{ where } y=e^{-\beta E}\Gamma_{q}(E,\tilde{q})
\nonumber\\
&\text{with }\Gamma_{q}(x,z)=\prod_{k=0}^{\infty}\frac{1}{1-(1-q)q^k z x+(1-q)q^{2k}z^2}
\label{npf2}
\end{align}
Hence, the partition function $Z_n$ resembles the calculation of the $n$-th moment of $y$. \eqref{npf2} tells us that all $2n$ boundaries in this single connected geometry should have the same energy $E$, as in the case of JT gravity. Remark that $\Gamma_{q}(x,z)$ is the $q$-deformed coherent state vector for $q$-Hermite polynomials $H^{(q)}_{n}(x)$, then $Z_1$ can be considered as a $q$-analogue of the Segal-Bargmann transform\cite{Gbron}
\begin{align}
Z_1=\int_{-2/\sqrt{1-q}}^{2/\sqrt{1-q}} e^{-\beta E}\Gamma_q(E,\tilde{q})\nu_q(E)dE
\label{z1}
\end{align}
Note that equation \eqref{z1} is presented as the partition function of DSSYK model in the presence of the EOW brane in \cite{Okuyama2}.

$Z_n$ as described in \cite{Penington}, corresponds to the fully connected wormhole saddle point solution in the gravitational path integral of the PSSY model. This solution dominates after the Page time. To get the whole picture of the Page curve, it is necessary to sum over all saddle point configurations, including the fully disconnected one, which is given by $Z_1^n$. Then, to obtain similar results in the chord Hilbert space, a straightforward approach is to suggest a $q$-compound Poisson distribution \cite{Anshelevich} while taking into account the information provided by equation \eqref{z1}. This can be realized by non-commutative random variable as sums of creation, annihilation and gauge operators on a $q$-Fock space. This issue will be addressed in future work.

In summary, the adoption of chord Hilbert space terminology and its corresponding transfer matrix $T$, is believed to enable the recreation of the Euclidean wormhole structure observed in JT gravity. This is due to the fact that all the gluing and slicing in the gravitational path integral have analogous versions in chord Hilbert space. However, the physical significance of this, particularly how to transition from chord Hilbert space to the SYK model, remains missing.
\section*{Acknowledgements}
We would like to thank Chushun Tian for the interesting disscussion and useful suggestions. We also acknowledge discussions with Hanteng Wang. This work is supported by the National Natural Science Foundation of China Grant No. 11925507 and No. 12047503.

\section*{Data availability statements}
Any data that support the findings of this study are included within the article.
\begin{appendices}
\renewcommand{\thesection}{\Alph{section}}
\numberwithin{equation}{section}
\section{Free probability: additive  convolution}
\label{appd:1}
The following has been adapted from \cite{Biane2}.

Considering two probability measure $\mu_a$ and $\mu_b$ on $\mathbb{R}$, the free additive convolution between them is denoted by $$\mu_{a+b}=\mu_a\boxplus\mu_b$$
The resolvent is defined by
\begin{align*}
G_{\mu}(z)=\int\frac{d\mu (x)}{z-x}=\frac{1}{z}M_{\mu}(1/z)
\end{align*}
where $M_{\mu}(z)=\sum_{n\geq 0}m_n z^n$ is the generating function of the moments of $\mu$. 
Let $K_{\mu}(z)$ be its inverse function and 
\begin{align*}
K_{\mu}(z)=R_{\mu}(z)+\frac{1}{z}
\end{align*}
$R_{\mu}(z)$ here is called $\mathcal{R}$-transform of $\mu$, it has a power series form in terms of free cumulants
\begin{align*}
R_{\mu}(z):=\sum_{n=0}k_{n+1}z^n
\end{align*}
with 
\begin{align*}
m_n=\sum_{\pi\in NC(n)}\prod_ {B_i\in\pi}k_{\vert B_i\vert}
\end{align*}
this is equivalent to 
\begin{align*}
M_{\mu}\left(\frac{z}{1+zR(z)}\right)=1+zR(z)
\end{align*}
Therefore 
\begin{align*}
G_{\mu}(R_{\mu}(z)+1/z)=z
\end{align*}
we recover the definition of $K_{\mu}(z)$.
This $\mathcal{R}$-transform  is additive
\begin{align*}
R_{\mu_{a+b}}(z)=R_{\mu_{a}}(z)+R_{\mu_{b}}(z)
\end{align*}
Thus,
\begin{align*}
K_{\mu_{a+b}}(z)=K_{\mu_{a}}(z)+R_{\mu_{b}}(z)=R_{\mu_{a}}(z)+K_{\mu_{b}}(z)
\end{align*}
Define
\begin{align*}
H_{\mu_a}(z):=K_{\mu_{a+b}}(G_{\mu_a}(z))=z+R_{\mu_b}(G_{\mu_a}(z))
\end{align*}
and we denote its inverse function $\omega_{\mu_a}(z)$, then 
\begin{align*}
G_{\mu_{a+b}}(H_{\mu_a}(z))=G_{\mu_a}(z)\text{ $\Leftrightarrow$ }G_{\mu_{a+b}}(H_{\mu_a}(\omega_{\mu_a}(z)))=G_{\mu_a}(\omega_{\mu_a}(z))
\end{align*}
therefore we get
\begin{align*}
G_{\mu_{a+b}}(z)=G_{\mu_a}(\omega_{\mu_a}(z))=G_{\mu_b}(\omega_{\mu_b}(z))
\end{align*}
we say that the resolvent  $G_{\mu_{a+b}}(z)$ is subordinated to $G_{\mu_a}(z)$ (or $G_{\mu_a}(z)$) via $\omega_{\mu_a}(z)$ (or $\omega_{\mu_b}(z)$ respectively). $\omega_{\mu_a}(z)$ and $\omega_{\mu_b}(z)$ are referred to as the subordination functions. The support of $\mu_{a+b}$ is then controlled by these functions. In order to see this, we define a function $\psi_{\mu_a}(u)$ which is a map $\mathbb{R}\rightarrow\mathbb{R}$ and  satisfying
\begin{align*}
\psi_{\mu_a}(u)=H_{\mu_a}(u+i v(u))=\text{Re} H_{\mu_a}(u+i v(u))
\end{align*}
$\text{Re}$ means the real part. There are two possibilities for the domain of $\psi_{\mu_a}(u)$. If $u$ is outside of the support of $\mu_a$, $H_{\mu_a}(u)$ is automatically real. If $u$ is inside of the support of $\mu_a$, we need find a function $v(u)$ such that Im$H_{\mu_a}(u+iv(u))=0$.
We can get the density of $\mu_{a+b}$ via Stieltjes inversion formula
\begin{align*}
\rho_{\mu_{a+b}}(x)=\frac{-1}{\pi}\lim_{y\rightarrow 0}\text{Im }G_{\mu_{a+b}}(x+iy)=\frac{-1}{\pi}\lim_{y\rightarrow 0}\text{Im }G_{\mu_{a}}(\omega_{\mu_a}(x+iy))
\end{align*}
Im means the imaginary part. At the point $\psi_{\mu_a}(u)$, the measure $\mu_{a+b}$ has a density
\begin{align*}
\rho_{\mu_{a+b}}(\psi_{\mu_a}(u))=\frac{-1}{\pi}\text{Im}G_{\mu_{a}}(u+i v(u))
\end{align*} 
$\psi_{\mu_a}(u)$ controls the support of $\mu_{a+b}$. 

Now consider 
\begin{align*}
\mu_a=(1-r)\delta_0+r\delta_{\theta}
\end{align*}
with $0\leq r\leq 1$ and $\theta\in\mathbb{R}^{+}$. If $r\ll 1$, $\mu_a\thicksim \delta_0$, then $G_{\mu_a}(z)=1/z$ and we get
\begin{align*}
H_{\mu_a}(z)=z+R_{\mu_b}(1/z)=K_{\mu_b}(1/z)\Leftrightarrow \omega_{\mu_a}(z)=1/G_{\mu_b}(z)
\end{align*}
Therefore $\psi_{\mu_a}(u)$ is inside the support of $\mu_b$ if $u$ is inside the support of $\mu_a$.
Moreover $H_{\mu_a}(\theta)=E_{\text{out}}$ gives an outlier of the support of $\mu_{b}$ since it is derived by 
\begin{align*}
\omega_{\mu_a}(E_{\text{out}})=1/G_{\mu_b}(E_{\text{out}})=\theta
\text{ with }\theta>1/G_{\mu_b}(E_{\text{max}}) \text{ where } E_{\text{max}}=\max[\text{supp}\mu_b]
\end{align*}
This means whenever we have a spike eigenvalue outside $\mu_a$, it creates outlier  of $\mu_{a+b}$ which is outside the support of  $\mu_{b}$.
The above equation coincides with eq $14$ in \cite{Wu}, which identifies the location of the separated eigenvalue in the spectrum of \eqref{hamiltonian} with $c=1$. This means that if $c$ is small enough, we can use free probability theory to explain the phase transition in the spectrum of \eqref{hamiltonian}.

Now if we consider 
$\mu_b$ be a semi-circular distribution:
\begin{align*}
\rho_{\mu_b}(x)=\frac{1}{2\pi}\sqrt{4-x^2}\text{ with }G_{\mu_b}(z)=\frac{z-\sqrt{z^2-4}}{2}
\end{align*}
One has then
\begin{align*}
R_{\mu_b}(z)=z
\end{align*}
With $\mu_a$ being an arbitrary probability measure, we get
\begin{align*}
H_{\mu_{a}}(z)=z+G_{\mu_a}(z)
\end{align*}
the support of $\mu_{a+b}$ is given by
\begin{align*}
\psi_{\mu_a}(u)=u+\int\frac{(u-x)d\mu_a(x)}{(u-x)^2+v^2(u)}
\end{align*}
with $v(u)$ satisfying
\begin{align*}
\int\frac{d\mu_a(x)}{(u-x)^2+v^2(u)}=1
\end{align*}
The measure $\mu_{a+b}$ 's density at $\psi_{\mu_a}(u)$ is then given by
\begin{align*}
\rho_{\mu_{a+b}}(\psi_{\mu_a}(u))=\frac{-1}{\pi}\text{Im}G_{\mu_{a}}(u+i v(u))=\frac{v(u)}{\pi}
\end{align*}
It is easy to see that if the support of $\mu_a$ has multiple intervals, the solution of $v(u)$ could be in multiple intervals.
\section{Averaged Weight factor $\tilde{q}$ in an intersection $D_c\Psi_{\alpha}D_c\Psi_{\alpha}$}
\label{appd:2}
Given $c=2^{N/2-k}$, we decompose $D_c$ into $k$ parts:
$$
D_c=\frac{1}{2^k} \left(m_0+m_1+\cdots +m_{k-1}\right)
$$
each $m_j$ with $0\leq j\leq k-1$ are linear combinations of products of Majorana fermions: 
\begin{align}
m_j&:=m_{j}(k)\nonumber\\
&=(-i)^j\sum_{l_1=0}^{k-2}\sum_{l_2=l_1+1}^{k-2}\dots\sum_{l_j=l_{j-1}+1}^{k-2}\prod_{n=0}^{j-1}\psi_{N-2l_{j-n}-1}\psi_{N-2l_{j-n}}(\mathbb{1}_{2^{N/2}}+(-i)^{N/2}\prod_{r=1}^N\psi_{r})
\end{align}
Therefore they have their own commutative relations with $\Psi_{\alpha}$. Remembering that $\Psi_{\alpha}$ is a product of $p$ Majorana fermions defined in \eqref{gammarel}, we have
\begin{align}
\prod_{n=0}^{j-1}\psi_{N-2l_{j-n}-1}\psi_{N-2l_{j-n}}\Psi_{\alpha}=(-1)^{p+l}\Psi_{\alpha}\prod_{n=0}^{j-1}\psi_{N-2l_{j-n}-1}\psi_{N-2l_{j-n}}
\end{align}
with a given index set $\alpha$ and a fixed ascending ordered set $\lbrace l_1,l_2,\cdots,l_{j-1},l_j\rbrace$. $l$ is the number of elements in common between these two index sets. By summing the index set $\alpha$, the contraction between two $\Psi_{\alpha}$ intersecting with a $m_j$ is given by
\begin{align*}
\binom{N}{p}^{-1}\sum_{\alpha}\Psi_{\alpha} m_j\Psi_{\alpha}=\binom{N}{p}^{-1}\sum_{l=0}^{\min[2j,N-2j]}\binom{2j}{l}\binom{N-2j}{p-l}(-1)^{p+l}m_j\mathbb{1}
\end{align*}
we drop $(-1)^p$ since we assume $p$ to be even.
Note that $\binom{N}{p}^{-1}\binom{2j}{l}\binom{N-2j}{p-l}$ is the probability that a subset $A\subset [N]$ of size $p$ has $l$ elements in common with another fixed subset $B\subset [N]$ of size $2j$.

We call $q_j$ the average factor to count the exchange between $m_j$ and $\Psi_{\alpha}$ and it is defined as
\begin{align}
q_j=\frac{1}{\binom{N}{p}}\sum_{l=0}^{\min[2j,N-2j]}\binom{2j}{l}\binom{N-2j}{p-l}(-1)^{l}
\label{q_j}
\end{align}

On the other hand, we know that both $D_c$ and its components $m_j$ are diagonal matrices, and have relations 
\begin{align*}
m_j D_c=\left\{\begin{array}{ll}2(k-1) D_c \quad\quad & 0<j<k-1 \\ 2 D_c \quad & j=0\quad\text{or}\quad j=k-1\end{array}\right.
\end{align*}
Combining the above information, we have
\begin{align}
\binom{N}{p}^{-1}\sum_{\alpha}\text{tr } D_c\Psi_{\alpha}D_c\Psi_{\alpha} &=\frac{1}{2^{k}}\text{tr}\Psi_{\alpha}(q_0 m_0+ q_1 m_1+\cdots+q_{k-1} m_{k-1})D_c\Psi_{\alpha}\nonumber\\
&=\frac{1}{2^{k-1}}(q_0 + (k-1)(q_1+\cdots+q_{k-2})+q_{k-1})\text{tr} D_c\Psi_{\alpha}\Psi_{\alpha}\nonumber\\
&=\tilde{q}\text{tr} D_c =\frac{\tilde{q}}{2^{k}}=\tilde{q}\text{tr} D^2_r\text{tr} \Psi^2_{\alpha}
\end{align}
with $$\tilde{q}=\frac{q_0+(k-1)(q_1+\cdots+q_{k-2})+q_{k-1}}{2^{k-1}}$$ 
\section{Extract $m_n$ from its generating function}
\label{appd:3}
Two generating function $M(z)=\sum_{k\geq0} m_n z^n$ and $B(z)=\sum_{k\geq0} b_n z^n$ satisfy the relation 
$$
M(z)=\langle 0\vert z\frac{d}{dz}\log\frac{1}{1-\theta z B(z)}\vert 0\rangle
$$ 
the coefficient $b_i=b_i(x_0)$ is a polynomial of $x_0$ with the highest degree $i$, the vacuum expectation is taking as the integral over the distribution of  $x_0$ of a form $b_i(x_0)=c_i x_0^i+c_{i-2}x_0^{i-2}+c_{i-4}x_0^{i-4}+\cdots$. Thus, the coefficient $m_n$ is obtained 
\begin{align*}
m_n&=n[z^n]\langle 0\vert\log\frac{1}{1-\theta z B(z)}\vert 0\rangle\\
&=n[z^n]\langle 0\vert\sum_{k\geq0}\frac{\left(1+b_1 z+b_2 z^2+b_3 z^3+\cdots \right)^k \theta^k z^k}{k}\vert 0 \rangle\\
&=n\langle 0\vert\sum_{j=0}^n \theta^j 
\sum_{\substack{k_1+k_2+k_3+\cdots=j \\ k_1+2k_2+3k_3+\cdots =n- j}}\frac{\binom{k_1+k_2+k_3+\cdots}{k_1,k_2,k_3,\cdots}}{k_1+k_2+k_3+\cdots}\prod_{i}b_i^{k_i}\vert 0\rangle
\end{align*}
we assume $x_0$ is $q$-Gaussian distributed, it only has even moments, by taking the expectation value of $x_0$ and taking account of the form of $b_i(x_0)$, one needs to restrict the second sum equals to a even number, we assume $k_1+2k_2+\cdots+ 2j k_{2j}=2j$, with $j\in[0,\lfloor n/2\rfloor]$, we get
\begin{align}
m_n&=n\langle 0\vert\sum_{j=0}^{\lfloor n/2\rfloor}\theta^{n-2l}\sum_{k_1+2k_2+\cdots +2jk_{2j} =2j}\frac{\binom{n-2j}{k_1,k_2,\cdots,k_{2l},n-2l-\sum_{i=1}^{2j}k_i}}{n-2l} \prod_{i=1}^{2j}b_i^{k_i}\vert 0\rangle\nonumber\\
&=n\langle 0\vert\sum_{j=0}^{\lfloor n/2\rfloor}\theta^{n-2j}\sum_{\substack{l=1 \\k_1+k_2+\cdots +k_{l} =2j}}^{\min[p-2j,2j]}\frac{\binom{n-2j}{l}}{n-2j} \prod_{i=1}^{l}b_{k_i}\vert 0\rangle\nonumber\\
&=n\sum_{j=0}^{\lfloor n/2\rfloor}\theta^{n-2j}\sum_{\substack{l=1 \\k_1+k_2+\cdots +k_{l} =2j}}^{\min[p-2j,2j]}\frac{\binom{n-2j}{l}}{n-2j} \langle 0\vert\prod_{i=1}^{l}b_{k_i}\vert 0\rangle
\label{c1}
\end{align}
with 
$$
b_i=\mathbb{E}[x_1^{i}\vert x_0]=\sum_{m=0}^{\lfloor i/2\rfloor}c_{m,i}\sqrt{\tilde{q}}^{i-2m}H_{i-2m}^{(q)}(x_0)
$$
therefore
\begin{align}
\langle 0\vert \prod_{i=1}^{l}b_{k_i}\vert 0\rangle&=\langle 0\vert \prod_{i=1}^{l}\sum_{m_i=0}^{\lfloor k_i/2\rfloor}c_{m_i,k_i}\sqrt{\tilde{q}}^{k_i-2m_i}H_{k_i-2m_i}^{(q)}(x_0)\vert 0\rangle\nonumber\\
&=(\prod_{i=1}^{l}\sum_{m_i=0}^{\lfloor k_i/2\rfloor}c_{m_i,k_i}\sqrt{\tilde{q}}^{k_i-2m_i})\langle 0\vert\prod_{i=1}^l H_{k_i-2m_i}^{(q)}(x_0)\vert 0\rangle
\label{c2}
\end{align}
The final step in calculating $m_n$ from above is to know that
\begin{align}
&\mathbb{E}\left[\prod_{j=1}^k H_{n_j}^{(q)}(x_0)\right]\quad\text{with}\quad x_0=\frac{2\cos\theta}{\sqrt{1-q}}\nonumber\\
&=\int_{0}^{\pi} \frac{d \theta}{2 \pi}\nu_q d\theta \prod_{j=1}^{k} H_{n_{j}}^{(q)}(x_0)=\sum_{n_{i j}} \prod_{i=1}^{k}\left(\begin{array}{c}n_{i} \\ n_{i 1}, \ldots, n_{i k}\end{array}\right)_{q} \prod_{1 \leq i<j \leq k}[n_{i j}]_q ! q^{B}
\label{c3}
\end{align}
see explanation under equation \eqref{mm} and COROLLARY 3.4 of \cite{Viennot}. Combing \eqref{c3} and \eqref{c2} into \eqref{c1}, we get \eqref{mm}.
\end{appendices}

\end{document}